\documentclass[pra,showpacs]{revtex4}

\usepackage{graphics}       
\usepackage{bm}         
\usepackage{amsmath}        
\usepackage{latexsym}
\begin{document}
\title{Unambiguous comparison of the states of multiple quantum systems}
\author{Anthony Chefles}
\affiliation{School of Physics, Astronomy and Mathematics,
University of Hertfordshire,
       Hatfield AL10 9AB, Hertfordshire, UK}
\author{Erika Andersson}
\affiliation{Department of Physics, University of Strathclyde,
Glasgow G4 0NG, UK}
\author{Igor Jex}
\affiliation{Department of Physics, FNSPE, Czech Technical
University Prague, Br\v ehov\'a 7, 115 19 Praha, Czech Republic}
\begin{abstract}
\vspace{0.5cm}

We consider $N$ quantum systems initially prepared in pure states
and address the problem of unambiguously comparing them. One may
ask whether or not all $N$ systems are in the same state.
Alternatively, one may ask whether or not the states of all $N$
systems are different.  We investigate the possibility of
unambiguously obtaining this kind of information. It is found that
some unambiguous comparison tasks are possible only when certain
linear independence conditions are satisfied.  We also obtain
measurement strategies for certain comparison tasks which are
optimal under a broad range of circumstances, in particular when
the states are completely unknown. Such strategies, which we call
universal comparison strategies, are found to have intriguing
connections with the problem of quantifying the distinguishability
of a set of quantum states and also with unresolved conjectures in
linear algebra. We finally investigate a potential generalisation
of unambiguous state comparison, which we term unambiguous overlap
filtering.

\end{abstract}
\pacs{03.65.Bz, 03.67.-a, 03.67.Hk}
\maketitle
\section{Introduction}
\renewcommand{\theequation}{1.\arabic{equation}}
\setcounter{equation}{0} \label{sec:1}

In classical physics, the state of a physical system is the
information which specifies the values of its dynamical variables.
There are no fundamental limitations on the precision with which
these variables can be measured simultaneously. So, in classical
physics the state of a physical system can, in principle, be
measured with arbitrarily high accuracy. This contrasts with the
situation in quantum mechanics. The state of a quantum mechanical
system, is, in the best case scenario, represented by a normalised
vector in the system's associated Hilbert space. The only state
vectors that can be completely distinguished from one another are
vectors belonging to a known, orthogonal set.  As a consequence,
it is impossible to measure the state of a quantum system if we do
not have prior information specifying an orthogonal set to which
it belongs.

This fact has inspired a great deal of research relating to the
limitations imposed by the quantum formalism on our ability to
obtain information about the state of a quantum system.  In the
investigation of this matter, different approaches relating to
correspondingly distinct scenarios have been developed
\cite{review}. For example, in quantum state discrimination, the
aim is to determine the state as well as possible, given that we
know that the system has been prepared in one of a finite number
of possibly non-orthogonal states, and with a knowledge of their a
priori probabilities. In quantum state estimation, we do not have
this information and we aim to determine, as well as possible, a
completely unknown state, although perhaps with multiple copies of
this state.

In this paper, we investigate the possibility of obtaining a
different kind of information about quantum states.  Here, we are
given $N$ quantum systems, which are physically of the same kind
and have identical Hilbert spaces.  Each of the $N$ systems has
the same set of possible states which, throughout, we assume to be
pure.   Our aim is to determine whether or not the actual states
of these systems are all identical, or all different. This kind of
procedure is known as quantum state comparison.

There has recently been a substantial amount of work done in
relation to this matter. The first investigation focused on
comparing the states of just two quantum systems \cite{BCJ}.
Subsequently, the possibility of comparing unitary operators was
proposed and explored \cite{ucomp}.  Also, the links between
comparison of pure quantum states and discrimination between
mixed quantum states are described in \cite{BCJ} and
\cite{Rudolph}. Two applications in quantum communications where
state comparison plays an important role are quantum
fingerprinting \cite{Finger} and quantum digital signatures
\cite{Signatures}. Quantum state comparison has also been proposed
as a technique for quantum computer stabilisation \cite{Sym}.

More recently, we have explored numerous aspects of the problem of
comparing the states of many quantum systems \cite{JAC}.  The
present article is in many ways a companion article to this work.
In \cite{JAC}, different kinds of comparison strategy were
examined for several sets of possible states, and techniques for
experimentally implementing these strategies were proposed. The
aim of the present paper is to present a detailed analysis of the
problem of unambiguous state comparison for multiple quantum
systems with multiple possible states.  We also give rigorous
proofs of some of the assertions made in \cite{JAC}.

In unambiguous state comparison, the aim is to establish the
identicality of, or differences among the states of many quantum
systems, with zero probability of error. This cannot be achieved
in a completely reliable way if the states are non-orthogonal.  We
must, in general, allow for a non-zero probability that our
comparison attempt will yield an inconclusive result.

We begin in section \ref{sec:2} with a brief review of generalised
quantum measurements, which we shall use extensively throughout
this article. For a more detailed treatment of such matters, see
\cite{Kraus}.  Our investigation of unambiguous state comparison
commences in section \ref{sec:3}. Here, we examine the problem of
confirming, unambiguously, that all $N$ particles have been
prepared in the same state, a task that we term identicality
confirmation.  We show that if this is to succeed with non-zero
success probability for each element of a discrete set of
$N$-particle states, then this set, or equivalently, the set of
possible single-particle states, must be linearly independent.

On the other hand, if we wish to unambiguously confirm that the
$N$ particles are not all prepared in the same state, then no such
restriction applies.  In section \ref{sec:4} we describe a
universal measurement strategy for unambiguous confirmation that
not all $N$ particles have been prepared in the same state.  This
strategy is optimal under a broad range of circumstances, in
particular when the single-particle states are
uniformly-distributed.

Alternatively, we might wish to know when the states of the $N$
particles are all different from each other.  We address this
problem in section \ref{sec:5}.  Here we show that the possible
single-particle states must obey a weaker linear independence
condition than that in section \ref{sec:3}. We find that this
linear independence condition is always sufficient and obtain a
universal measurement for confirming that all $N$ particles are in
different states. This measurement succeeds with non-zero
probability whenever this linear independence condition is
satisfied.  It is also shown to be optimal when all $N$-particle
product states are possible.

In section \ref{sec:6}, we examine the above universal unambiguous
comparison strategies in greater detail.  In particular, we
investigate the possibility that their associated success
probabilities are suitable measures of the distinguishability of
the states of the individual particles, considered as possible
states of a single particle. For this to be the case, these
probabilities must satisfy certain conditions. In particular, they
must be non-increasing under all deterministic quantum operations
which transform these pure states into another set of pure states.
We find that this is indeed the case for the second of our
universal comparison strategies. The question of whether or not it
is for the first strategy is found to be equivalent to a certain,
currently unresolved conjecture in linear algebra.

The final topic we discuss, in section \ref{sec:7}, is a procedure
we term unambiguous overlap filtering.  In unambiguous state
comparison, our aim is to determine, with zero probability of
error, whether or not a number of quantum systems have been
prepared in the same pure state. This is equivalent to determining
whether or not they have mutual overlap equal to 1.  A potential
generalisation of this procedure is the unambiguous determination
of whether or not they have overlap equal to some fixed
${\omega}{\in}[0,1]$. We concentrate on the states of just a pair
of systems.  We find that, when these states are pure but
otherwise arbitrary and unknown, then it is impossible, with any
non-zero probability, and for any value of ${\omega}{\in}[0,1]$,
to unambiguously confirm that they have overlap equal to
${\omega}$.  We also find that it is impossible to unambiguously
confirm, with non-zero probability, that their overlap is not
equal to ${\omega}$ unless ${\omega}=1$, which corresponds to
difference detection. Therefore, the only possible kind of
unambiguous overlap filtering for this set of states is
unambiguous difference detection.  We conclude in section
\ref{sec:8} with a general discussion of our results.

\section{State comparison and generalised measurements}
\renewcommand{\theequation}{2.\arabic{equation}}
\setcounter{equation}{0} \label{sec:2} In this paper, we will
investigate problems relating to quantum state comparison.  We
will be focusing on questions such as `are the states of $N$
quantum systems all identical?' and `are the states of $N$ quantum
systems all different?'.  If we know the states of the individual
systems, then questions such as these can be readily answered by
comparing explicit expressions for these states.  If, however, we
do not know these states in advance, then to obtain the required
information about our $N$-particle system, we must perform
measurements on it. In order to assess the scope and limitations
of the quantum measurement process with respect to such problems,
we should consider the most general class of measurements that are
physically possible on our $N$-particle system. In introductory
texts on quantum mechanics, e.g. \cite{BJ}, the measurements which
are most commonly considered are von Neumann measurements on a
single quantum system. There are two possible extensions of this
class of measurements, which, when permitted, make accessible to
us the full range of physically possible measurement procedures on
an $N$-particle system. The first extension involves lifting the
restriction to measurements on single particles, and to allow for
the possibility of collective measurements on several, potentially
all $N$ particles. In numerous contexts in quantum information
theory, it has been found that such collective measurements can be
used to achieve tasks that are impossible using single-particle
measurements.  Perhaps the best-known example is the Bell
measurement, which is an essential component of many important
quantum communications protocols, such as teleportation
\cite{Teleport} and dense coding \cite{Coding}.

The second extension is the lifting of the restriction to von
Neumann measurements.  A much broader class of measurements on
quantum systems are possible, which are appropriately known as
generalised measurements.  The statistical properties of such a
measurement are described in terms of a set of positive
 operators \cite{Footnote1} known as positive, operator-valued
measure (POVM) elements. If we have a generalised measurement
(henceforth simply measurement) with $K$ possible outcomes,
indexed by $k=1,{\ldots},K$, then associated with the $k$th
outcome is a POVM element $E_{k}$. These operators act on the
Hilbert space of the total system under investigation and we
denote this by ${\cal H}_{tot}$.

The main purpose served by the POVM elements is to provide the
probability distribution for the measurement outcomes for each
initial state. Given that the initial state is described by a
density operator ${\rho}$ acting on ${\cal H}_{tot}$, the
probability of obtaining outcome $k$ for this measurement will be
denoted by $P_{k}({\rho})$. Whether or not a generalised
measurement is possible with specified probabilities
$P_{k}({\rho})$ depends on whether or not the following two
necessary and sufficient conditions are satisfied. Firstly, there
must exist a set of $K$ positive operators $E_{k}$, which will be
the POVM elements of the measurement, such that the probabilities
have the form
\begin{equation}
P_{k}({\rho})=\mathrm{Tr}({\rho}E_{k}).
\end{equation}
When the system is initially in a pure state
${\rho}=|{\Psi}{\rangle}{\langle}{\Psi}|$, this expression may be
written as
\begin{equation}
P_{k}({\Psi})={\langle}{\Psi}|E_{k}|{\Psi}{\rangle}.
\end{equation}
Secondly, the $K$ POVM elements must form a resolution of the
identity,
\begin{equation}
\sum_{k=1}^{K}E_{k}=1_{tot},
\end{equation}
where $1_{tot}$ is the identity operator on the Hilbert space
${\cal H}_{tot}$.  This expression is equivalent to the
normalisation of the outcome probability distribution for the all
initial states ${\rho}$,
\begin{equation}
\sum_{k=1}^{K}P_{k}({\rho})=1.
\end{equation}
A collective, generalised measurement is the most general
measurement procedure that can be carried out on a multiparticle
quantum system.  In our subsequent discussion of quantum state
comparison, we will seek to obtain results of maximum possible
generality.  With this in mind, the properties of collective,
generalised measurements, in particular, what can and what cannot
be achieved with them, will be of considerable importance.

\section{Unambiguous identicality confirmation for linearly independent states}
\renewcommand{\theequation}{3.\arabic{equation}}
\setcounter{equation}{0} \label{sec:3} Consider the following
situation.  We have $N$ particles indexed by the label
$j=1,{\ldots},N$. Associated with each of them is a copy of the
finite dimensional Hilbert space ${\cal H}$.  We shall denote the
dimensionality of this space by $D({\cal H})$.  The Hilbert space
of the entire $N$-particle system is ${\cal H}_{tot}={\cal
H}^{{\otimes}N}$.  Each particle has been prepared in an element
of the set of $M$ pure states $\{|{\psi}_{\mu}{\rangle}\}$, with
${\mu}=1,{\ldots},M$. All of the states in this set are different
from each other. Throughout this article, two pure states, say
$|{\psi}_{1}{\rangle}$ and $|{\psi}_{2}{\rangle}$, will be
considered identical if
$|{\langle}{\psi}_{1}|{\psi}_{2}{\rangle}|=1$ and different
otherwise.  We denote the actual state of particle $j$ by
$|{\psi}_{{\mu}_{j}}{\rangle}$. We may write the state of the
$N$-particle system in tensor product form,
\begin{equation}
|{\Psi}_{{\mu}_{1}{\ldots}{\mu}_{N}}{\rangle}=\bigotimes\limits_{j=1}^{N}|{\psi}_{{\mu}_{j}}{\rangle}.
\end{equation}
Our $N$-particle system has been prepared in one of the $M^{N}$
states $|{\Psi}_{{\mu}_{1}{\ldots}{\mu}_{N}}{\rangle}$. We would
like to find a measurement that enables us to unambiguously
confirm when all $N$ particles have been prepared in the same
state.

This task can sometimes be achieved by unambiguous state
discrimination.  It is known that one can determine unambiguously
in which element of a finite set of pure states a quantum system
has been prepared, with non-zero probability for each element, if
and only if this set is linearly independent \cite{Chefles1}. This
procedure could clearly be applied to the problem at hand for
linearly independent
$|{\Psi}_{{\mu}_{1}{\ldots}{\mu}_{N}}{\rangle}$.  If this linear
independence condition is satisfied, then we could unambiguously
determine, with non-zero probability, in which of the states
$|{\Psi}_{{\mu}_{1}{\ldots}{\mu}_{N}}{\rangle}$ our $N$-particle
system has been prepared. When this unambiguous discrimination
attempt succeeds, it will give the values of the ${\mu}_{j}$. We
need then only compare the values of the ${\mu}_{j}$ and determine
whether or not they are all equal to accomplish the state
comparison task in question. So, we see that linear independence
of the $|{\Psi}_{{\mu}_{1}{\ldots}{\mu}_{N}}{\rangle}$ is a
sufficient condition for being able to unambiguously confirm that
all $N$ particles have been prepared in the same state.

Furthermore, one can easily show that the linear independence of
the possible $N$-particle states
$|{\Psi}_{{\mu}_{1}{\ldots}{\mu}_{N}}{\rangle}$ is equivalent to
that of the possible single-particle states
$|{\psi}_{\mu}{\rangle}$.  One way to prove this is to make use of
the fact that, for any density operator ${\rho}$ which is a
statistical mixture of $M({\rho})$ pure states, these states are
linearly independent if and only if
$\mathrm{rank}({\rho})=M({\rho})$. With this in mind, consider the
density operators
\begin{eqnarray}
{\rho}_{1}&=&\frac{1}{M}\sum_{{\mu}=1}^{M}|{\psi}_{\mu}{\rangle}{\langle}{\psi}_{\mu}|,
\\
{\rho}_{N}&=&\frac{1}{M^{N}}\sum_{{\mu}_{1},{\ldots},{\mu}_{N}=1}^{M}|{\Psi}_{{\mu}_{1}{\ldots}{\mu}_{N}}{\rangle}{\langle}{\Psi}_{{\mu}_{1}{\ldots}{\mu}_{N}}|={\rho}_{1}^{{\otimes}N}.
\end{eqnarray}
Clearly, $M({\rho}_{1})=M$ and $M({\rho}_{N})=M^{N}$.  From the
second part of Eq. (3.3), we see that
$\mathrm{rank}({\rho}_{N})=[\mathrm{rank}({\rho}_{1})]^{N}$. From
these observations, we find that
$\mathrm{rank}({\rho}_{N})=M({\rho}_{N})$ if and only if
$\mathrm{rank}({\rho}_{1})=M({\rho}_{1})$ and this completes the
proof.  We can then conclude that a sufficient condition for being
able to unambiguously confirm, with non-zero probability,
identicality of the states of the $N$ particles is that the set of
possible single-particle states is linearly independent.

We shall see here that linear independence of the possible
single-particle states or, equivalently, that of the possible
$N$-particle states, is also a necessary condition for unambiguous
confirmation that all $N$ particles are in the same state.  To see
this, let us formulate the problem as a generalised measurement in
the way described in section \ref{sec:2}.  The measurement will
have three potential outcomes: `same', `different' and `?'.  The
first pair of outcomes signal that all $N$ particles have been
prepared in the same state, and that they have not all been
prepared in the same state respectively.  The third possible
outcome, `?', is the inconclusive result.  On obtaining this, we
are none the wiser about whether or not the states of all $N$
particles are identical. Corresponding to these possible outcomes
are the POVM elements $E_{same}, E_{diff}$ and $E_{?}$.  These are
positive operators acting on ${\cal H}_{tot}$ which satisfy the
resolution of the identity in Eq. (2.3).

The measurement outcome that we are particularly interested in is that which signals that all $N$ particles have been prepared in the
same state.   We denote by $P_{same}({\Psi}_{{\mu}_{1}{\ldots}{\mu}_{N}})$ the probability of obtaining this `same' result for the
initial state $|{\Psi}_{{\mu}_{1}{\ldots}{\mu}_{N}}{\rangle}$.  From Eq. (2.2), we see that this is given by
\begin{equation}
P_{same}({\Psi}_{{\mu}_{1}{\ldots}{\mu}_{N}})={\langle}{\Psi}_{{\mu}_{1}{\ldots}{\mu}_{N}}|E_{same}|{\Psi}_{{\mu}_{1}{\ldots}{\mu}_{N}}{\rangle}.
\end{equation}
For the measurement result to be unambiguous and to succeed with
non-zero probability whenever all $N$ particles have been prepared
in the same state, we require that
$P_{same}({\Psi}_{{\mu}_{1}{\ldots}{\mu}_{N}})>0$ if and only if
the ${\mu}_{j}$ are all equal.

To prove that this requirement can only be satisfied if the
single-particle states are linearly independent, let us consider
the $N$-particle states of the form
$|{\Psi}_{{\mu}_{1}{\mu}{\ldots}{\mu}}{\rangle}$.  Here, particle
1 is in the state $|{\psi}_{{\mu}_{1}}{\rangle}$ and the remaining
particles are all in the state $|{\psi}_{\mu}{\rangle}$.  It is
clear from the above discussion that the result `same' can be
obtained with non-zero probability if and only if
${\mu}_{1}={\mu}$.  We may then write
\begin{equation}
P_{same}({\Psi}_{{\mu}_{1}{\mu}{\ldots}{\mu}})=p_{{\mu}}{\delta}_{{\mu}_{1}{\mu}},
\end{equation}
for some $p_{{\mu}}>0$.  Let us now attempt to write the set of possible states of particle 1 as superpositions of each other,
\begin{equation}
|{\psi}_{{\mu}_{1}}{\rangle}=\sum_{{\nu}=1}^{M}f_{{\mu}_{1}{\nu}}|{\psi}_{\nu}{\rangle}.
\end{equation}
The $|{\psi}_{{\mu}_{1}}{\rangle}$ are linearly independent if and
only if the only possible values of the coefficients
$f_{{\mu}{\nu}}$ are $f_{{\mu}{\nu}}={\delta}_{{\mu}{\nu}}$. In
fact, this is equivalent to
$|f_{{\mu}{\nu}}|={\delta}_{{\mu}{\nu}}$, a formulation of this
condition that will be more convenient.

Applying Eq. (3.6) to the $N$-particle state
$|{\Psi}_{{\mu}_{1}{\mu}{\ldots}{\mu}}{\rangle}$, we obtain
\begin{equation}
|{\Psi}_{{\mu}_{1}{\mu}{\ldots}{\mu}}{\rangle}=\sum_{{\nu}=1}^{M}f_{{\mu}_{1}{\nu}}|{\Psi}_{{\nu}{\mu}{\ldots}{\mu}}{\rangle},
\end{equation}
which gives us
\begin{equation}
{\langle}{\Psi}_{{\mu}_{1}{\mu}{\ldots}{\mu}}|E_{same}|{\Psi}_{{\mu}_{1}{\mu}{\ldots}{\mu}}{\rangle}=\sum_{{\nu},{\nu}'=1}^{M}f^{*}_{{\mu}_{1}{{\nu}'}}f_{{\mu}_{1}{{\nu}}}{\langle}{\Psi}_{{\nu}'{\mu}{\ldots}{\mu}}|E_{same}|{\Psi}_{{\nu}{\mu}{\ldots}{\mu}}{\rangle}.
\end{equation}
From the Cauchy-Schwarz inequality, we see that
\begin{eqnarray}
|{\langle}{\Psi}_{{\nu}'{\mu}{\ldots}{\mu}}|E_{same}|{\Psi}_{{\nu}{\mu}{\ldots}{\mu}}{\rangle}|^{2}&{\leq}&{\langle}{\Psi}_{{\nu}'{\mu}{\ldots}{\mu}}|E_{same}|{\Psi}_{{\nu}'{\mu}{\ldots}{\mu}}{\rangle}{\langle}{\Psi}_{{\nu}{\mu}{\ldots}{\mu}}|E_{same}|{\Psi}_{{\nu}{\mu}{\ldots}{\mu}}{\rangle}
\nonumber
\\
&=&p_{{\nu}'}p_{{\nu}}{\delta}_{{\nu}'{\mu}}{\delta}_{{\nu}{\mu}},
\end{eqnarray}
from which we obtain
\begin{equation}
{\langle}{\Psi}_{{\nu}'{\mu}{\ldots}{\mu}}|E_{same}|{\Psi}_{{\nu}{\mu}{\ldots}{\mu}}{\rangle}=p_{{\mu}}{\delta}_{{\nu}'{\mu}}{\delta}_{{\nu}{\mu}}.
\end{equation}
Substitution of this into Eq. (3.8) and making use of Eqs. (3.4)
and (3.5) gives
\begin{equation}
|f_{{\mu}_{1}{\mu}}|^{2}p_{\mu}=p_{\mu}{\delta}_{{\mu}_{1}{\mu}}.
\end{equation}
All of the probabilities $p_{\mu}$ are greater than zero, so we
finally obtain
\begin{equation}
|f_{{\mu}_{1}{\mu}}|={\delta}_{{\mu}_{1}{\mu}},
\end{equation}
showing that the states must be linearly independent.  This
completes the proof.

So, we see that a necessary and sufficient condition for
confirming identicality of the states of all $N$ particles is a
condition which is both necessary and sufficient for unambiguously
discriminating among all possible $N$-particle states, or
equivalently, all possible single-particle states.

Unambiguous state comparison may be viewed as unambiguous
discrimination between two subsets of the entire set of possible
$N$-particle pure states. In one subset, all $N$ particles are
prepared in the same state, and in the other, they are not.  The
fact that unambiguous discrimination between these two subsets is
possible if and only if we can unambiguously discriminate among
all possible $N$-particle states is interesting and non-trivial.
Even though unambiguous discrimination among all individual states
in a set is possible only if the states are linearly independent,
Sun et al \cite{Sun} showed that  this condition is not always
necessary if we only wish to unambiguously discriminate among
subsets instead. However, it is necessary here.

\section{Universal unambiguous detection of at least one difference}
\renewcommand{\theequation}{4.\arabic{equation}}
\setcounter{equation}{0} \label{sec:4}
\subsection{Universal measurement strategy}

The preceding discussion showed that it is impossible to confirm,
unambiguously, that all $N$ particles have been prepared in the
same state, with non-zero success probability whenever they are
identical, unless the possible single-particle states are linearly
independent. However, there is the possibility that this
restriction does not apply to unambiguous confirmation of when
this is not the case, i.e. of when the states of at least two
particles are different. In this section, we shall describe a
measurement that enables us to achieve this. For all $N$-particle
states where the individual particles are not all in the same
state, our measurement unambiguously confirms this fact with
non-zero probability. The fact that it does this for all such
states allows to call it a universal measurement. We will describe
its key properties and then prove that under a large set of
circumstances, it is actually optimal. This is to say that it
gives the minimum average probability of giving an inconclusive
result. One particularly important set of circumstances under
which it is optimal is when the possible states of each of the $N$
systems are uniformly-distributed, that is, they are completely
arbitrary and unknown.

Without loss of generality, we may restrict our measurement to
have only two possible outcomes, `different' and `?', having POVM
elements $E_{diff}$ and $E_{?}$ respectively.  Any POVM element
$E_{same}$ indicating that all particles are in the same state can
simply be added to $E_{?}$ resulting in a coarse-graining of these
two outcomes.  This will also apply to the analysis in section
\ref{sec:5}.

Let the $N$ particles be prepared in arbitrary pure states in
${\cal H}$. We shall denote the state of particle $j$ by
$|{\psi}_{j}{\rangle}$. The state vector of the entire
$N$-particle system can then be any product state in ${\cal
H}_{tot}$. We write it as
\begin{equation}
|{\Psi}{\rangle}=\bigotimes\limits_{j=1}^{N}|{\psi}_{j}{\rangle}.
\end{equation}
From Eq. (2.2), we see that, for the $N$-particle state
$|{\Psi}{\rangle}$, the probability of obtaining conclusive
difference detection is
\begin{equation}
P_{diff}({\Psi})={\langle}{\Psi}|E_{diff}|{\Psi}{\rangle}.
\end{equation}
For any $N$-particle state $|{\Psi}{\rangle}$ which is of the form
$|{\Psi}{\rangle}=|{\psi}{\rangle}^{{\otimes}N}$ for some
single-particle state $|{\psi}{\rangle}{\in}{\cal H}$, we see that
all $N$ particles are in the same state.  If this state has
non-zero probability of being prepared, then the condition of
unambiguity implies that $P_{diff}({\Psi})=0$ for such a state.
However, we require that $P_{diff}({\Psi})>0$ for every
$N$-particle product state $|{\Psi}{\rangle}$ which is not of this
form.

To describe our measurement which accomplishes this task, it will
be necessary to decompose the total Hilbert space ${\cal H}_{tot}$
into two subspaces. These are the symmetric subspace and its
orthogonal complement, which we shall refer to as the asymmetric
subspace. We denote these by ${\cal H}_{sym}$ and ${\cal
H}_{asym}$ respectively. Let $P({\cal H}_{sym})$ and $P({\cal
H}_{asym})$ be the projectors onto ${\cal H}_{sym}$ and ${\cal
H}_{asym}$ respectively.  The POVM elements corresponding to
conclusive detection of at least one difference and the
inconclusive result are
\begin{eqnarray}
E_{diff}&=&P({\cal H}_{asym}), \\ E_{?}&=&P({\cal H}_{sym}).
\end{eqnarray}
We will now proceed to show that this measurement has the
properties described above.

There are two things that we must prove.  Firstly, we must prove
that if all $N$ particles are in the same state, i.e. if
$|{\Psi}{\rangle}=|{\psi}{\rangle}^{{\otimes}N}$ for some
$|{\psi}{\rangle}{\in}{\cal H}$, then the probability of obtaining
a conclusive difference detection is zero.  Secondly, we must show
that for every product state $|{\Psi}{\rangle}$ which is not of
this form, the probability of obtaining a conclusive result is
non-zero.  Equivalently, the probability of obtaining an
inconclusive result is unity if and only if all particles are in
the same state.  We shall prove that our measurement satisfies
these requirements.

To do so, we will obtain an explicit expression for the
probability of obtaining an inconclusive result, and show that
this can only take the value of unity when all $N$ particles have
been prepared in the same state. This expression will also be
investigated in greater detail in section \ref{sec:6}.  To this
end, let us denote by $S(N)$ the symmetric (or permutation) group
of degree $N$. The symmetric tensor product of
$|{\psi}_{1}{\rangle},{\ldots},|{\psi}_{N}{\rangle}$ is defined as
\cite{Bhatia}
\begin{equation}
|{\psi}_{1}{\rangle}{\vee}{\dots}{\vee}|{\psi}_{N}{\rangle}=\frac{1}{\sqrt{N!}}\sum_{{\sigma}{\in}S(N)}|{\psi}_{{\sigma}(1)}{\rangle}{\otimes}{\dots}{\otimes}|{\psi}_{{\sigma}(N)}{\rangle}.
\end{equation}
For the state $|{\Psi}{\rangle}$ in Eq. (4.1), the projector onto
the symmetric subspace acts as follows \cite{Bhatia}:
\begin{eqnarray}
P({\cal H}_{sym})|{\Psi}{\rangle}&=&\frac{1}{\sqrt{N!}}|{\psi}_{1}{\rangle}{\vee}{\dots}{\vee}|{\psi}_{N}{\rangle} \nonumber \\
&=&\frac{1}{N !}\sum_{{\sigma}{\in}S(N)}|{\psi}_{{\sigma}(1)}{\rangle}{\otimes}{\dots}{\otimes}|{\psi}_{{\sigma}(N)}{\rangle}.
\end{eqnarray}
For the inconclusive result POVM element in Eq. (4.4), the
probability that this result is obtained for the state
$|{\Psi}{\rangle}$ in Eq. (4.1) is
\begin{eqnarray}
P_{?}({\Psi})&=&{\langle}{\Psi}|P({\cal H}_{sym})|{\Psi}{\rangle} \nonumber \\
&=&\frac{1}{N!}{\langle}{\psi}_{1}|{\otimes}{\dots}{\otimes}{\langle}{\psi}_{N}|\sum_{{\sigma}{\in}S(N)}|{\psi}_{{\sigma}(1)}{\rangle}{\otimes}{\dots}{\otimes}|{\psi}_{{\sigma}(N)}{\rangle}
\nonumber
\\&=&\frac{1}{N!}\sum_{{\sigma}{\in}S(N)}{\langle}{\psi}_{1}|{\psi}_{{\sigma}(1)}{\rangle}{\dots}{\langle}{\psi}_{N}|{\psi}_{{\sigma}(N)}{\rangle}.
\end{eqnarray}
Consider now the Gram matrix ${\Gamma}=({\gamma}_{ij})$ where
${\gamma}_{ij}={\langle}{\psi}_{i}|{\psi}_{j}{\rangle}$. From Eq.
(4.7) it follows that
\begin{equation}
P_{?}({\Psi})=\frac{1}{N!}\sum_{{\sigma}{\in}S(N)}{\gamma}_{1{\sigma}(1)}{\dots}{\gamma}_{N{\sigma}(N)}.
\end{equation}
If we now make use of the following definition of the permanent of
an $N{\times}N$ matrix $A=(a_{ij})$,
\begin{equation}
\mathrm{per}(A)=\sum_{{\sigma}{\in}S(N)}a_{1{\sigma}(1)}{\dots}a_{N{\sigma}(N)},
\end{equation}
we finally obtain
\begin{equation}
P_{?}({\Psi})=\frac{1}{N!}\mathrm{per}({\Gamma}).
\end{equation}
For the probability of obtaining an inconclusive result to be
unity, it is clear from Eq. (4.10) that we must have
$\mathrm{per}({\Gamma})=N!$.  Consider now the summation in Eq.
(4.8), which is the Gram matrix permanent.  It is a sum of $N!$
terms (since there are $N!$ possible permutations of $N$ objects),
each of which is bounded in absolute value by 1 (since they are
all products of inner products of normalised states).
$\mathrm{Per}({\Gamma})$ attains the value of $N!$ if and only if
each of the terms in this summation is equal to 1.  For this to be
true, it is necessary that
$|{\psi}_{j}{\rangle}=e^{i{\phi}_{j}}|{\psi}{\rangle}$, for some
$|{\psi}{\rangle}{\in}{\cal H}$ and some phases ${\phi}_{j}$, i.e.
the states of the $N$ particles are identical.  To prove that this
is also sufficient, we note that $\mathrm{per}({\Gamma})=N!$ if
${\phi}_{j}=0$. However, these phase factors do not influence the
value of $\mathrm{per}({\Gamma})$. The reason why is that they
accumulate to form an overall phase of $|{\Psi}{\rangle}$, which
disappears in the evaluation of the expectation value of $P({\cal
H}_{sym})$. Consequently, whenever all $N$ particles are in the
same state, regardless of possible phase factors, we have
$\mathrm{per}({\Gamma})=N!$.  This proves that the probability of
obtaining an inconclusive result is unity if and only if all
particles are in the same state.

\subsection{Proof of optimality}
We will now show that, under a large set of circumstances, the
measurement described above is optimal.   By optimal, we mean that
it attains the minimum average probability of obtaining an
inconclusive result. For any particular measurement strategy, to
determine the average probability of an inconclusive result, it is
necessary to know the a priori probabilities of the states.  The
most general situation that can be considered is where the set of
possible states is the set of all product states in the space
${\cal H}_{tot}$, that is, any state of the form
$|{\Psi}{\rangle}$ shown in Eq. (4.1).  The prior distribution of
these states is described by an a priori probability density
(henceforth simply density function) which we denote by
$q({\psi}_{1},{\ldots},{\psi}_{N})$. This function is non-negative
and normalised as
\begin{equation}
{\int}d{\sigma}({\psi}_{1}){\ldots}d{\sigma}({\psi}_{N})q({\psi}_{1},{\ldots},{\psi}_{N})=1,
\end{equation}
where each integration is with respect to the single-particle, unitarily-invariant Haar measure $d{\sigma}$.  This measure is
normalised, i.e. ${\int}d{\sigma}({\psi})=1$.  The total probability of obtaining an inconclusive result is simply the average of the
inconclusive result probability over the set of $N$-particle product states $|{\Psi}{\rangle}$ with respect to the density function
$q$, that is,
\begin{eqnarray}
\overline{P_{?}({\Psi})}&=&{\int}d{\sigma}({\psi}_{1}){\ldots}d{\sigma}({\psi}_{N})q({\psi}_{1},{\ldots},{\psi}_{N})P_{?}({\Psi})
\nonumber \\ &=&\mathrm{Tr}({\varrho}E_{?}),
\end{eqnarray}
where we have defined the $N$-particle density operator
\begin{equation}
{\varrho}={\int}d{\sigma}({\psi}_{1}){\ldots}d{\sigma}({\psi}_{N})q({\psi}_{1},{\ldots},{\psi}_{N})\left(\bigotimes\limits_{j=1}^{N}|{\psi}_{j}{\rangle}{\langle}{\psi}_{j}|\right).
\end{equation}
We will now consider the special case where
$q({\psi},{\ldots},{\psi})>0$ for all $|{\psi}{\rangle}{\in}{\cal
H}$.  That is, all product states of the form
$|{\Psi}{\rangle}=|{\psi}{\rangle}^{{\otimes}N}$, where all $N$
particles are in the same state, are possible. We will show that
when this condition is satisfied, the measurement described by the
POVM elements in Eqs. (4.3) and (4.4) attains the minimum average
probability of inconclusive results.  To show this, we will first
prove that if
$q({\psi},{\ldots},{\psi})>0\;\;{\forall}\;\;|{\psi}{\rangle}{\in}{\cal
H}$, then the condition of unambiguity is equivalent to the
support of the POVM element $E_{diff}$ being a subspace of ${\cal
H}_{asym}$ \cite{Footnote2}.  We will then show that, under
conditions of optimality, $E_{diff}$ may be taken to be the
projector onto ${\cal H}_{asym}$.

To prove the first of these assertions, if
$q({\psi},{\ldots},{\psi})>0\;\;{\forall}\;\;|{\psi}{\rangle}{\in}{\cal
H}$, then any state of the form
$|{\Psi}{\rangle}=|{\psi}{\rangle}^{{\otimes}N}$ can occur.  If
the result of the measurement is to be unambiguous, then we must
ensure that none of these states give rise to a conclusive result
indicating difference. Formally, this requirement can be expressed
as
\begin{equation}
\left({\langle}{\psi}|^{{\otimes}N}\right)E_{diff}\left(|{\psi}{\rangle}^{{\otimes}N}\right)=0,
\end{equation}
for all $|{\psi}{\rangle}{\in}{\cal H}$.  We now make use of the
fact that the projector onto the symmetric subspace may be written
as
\begin{equation}
P({\cal H}_{sym})=D({\cal H}_{sym}){\int}d{\sigma}({\psi})(|{\psi}{\rangle}{\langle}{\psi}|)^{{\otimes}N}.
\end{equation}
Here, $D({\cal H}_{sym})$ is the dimensionality of ${\cal
H}_{sym}$.  Its value is
\begin{equation}
D({\cal H}_{sym})={D({\cal H})+N-1 \choose N}.
\end{equation}
We now make use of Eqs. (4.14) and (4.15) and calculate
$\mathrm{Tr}[P({\cal H}_{sym})E_{diff}]$, finding that
\begin{equation}
\mathrm{Tr}[P({\cal H}_{sym})E_{diff}]=0.
\end{equation}
The trace of the product of two positive operators, such as we
have here, is zero if and only if each eigenvector of one of the
operators whose corresponding eigenvalue is non-zero is orthogonal
to all eigenvectors of the other operator with non-zero
eigenvalues. This implies that the eigenvectors of $E_{diff}$ with
non-zero eigenvalues must be orthogonal to ${\cal H}_{sym}$.  So,
these eigenvectors must lie in the orthogonal complement of ${\cal
H}_{sym}$, which is ${\cal H}_{asym}$. This proves our first
assertion.

To prove the second, we make use of the fact that, as a consequence of the above argument, $E_{diff}$ can be spectrally decomposed in
terms of an orthonormal basis for ${\cal H}_{asym}$,
\begin{equation}
E_{diff}=\sum_{r=1}^{D({\cal H}_{asym})}e_{r}|e_{r}{\rangle}{\langle}e_{r}|,
\end{equation}
where $0{\leq}e_{r}{\leq}1$ and the $|e_{r}{\rangle}$ form an
orthonormal basis for ${\cal H}_{asym}$.  Also, $D({\cal
H}_{asym})$ is the dimensionality of ${\cal H}_{asym}$, which is
easily seen to be equal to $D({\cal H})^{N}-D({\cal H}_{sym})$. If
we now substitute this expression into Eq. (4.12) and make use of
the resolution of the identity in Eq. (2.3), we find that
\begin{eqnarray}
\overline{P_{?}({\Psi})}&=&\mathrm{Tr}({\varrho}E_{?})=1-\mathrm{Tr}({\varrho}E_{diff}) \nonumber \\
&=&1-\sum_{r=1}^{D({\cal H}_{asym})}e_{r}{\langle}e_{r}|{\varrho}|e_{r}{\rangle} \nonumber \\
&{\geq}&1-\sum_{r=1}^{D({\cal H}_{asym})}{\langle}e_{r}|{\varrho}|e_{r}{\rangle} \nonumber \\
&=&1-\mathrm{Tr}[{\varrho}P({\cal H}_{asym})].
\end{eqnarray}
That is, the minimum value of $\overline{P_{?}({\Psi})}$ is
obtained when $E_{diff}$ is the projector onto ${\cal H}_{asym}$.
From the resolution of the identity in Eq. (2.3), it follows that
the corresponding inconclusive result POVM element $E_{?}$ must be
equal to $P({\cal H}_{sym})$.  So, we have shown that, under the
specified conditions, the measurement described by the POVM
elements in Eqs. (4.3) and (4.4) is optimal.  Eq. (4.19) can be
written in the form
\begin{equation}
\overline{P_{?}({\Psi})}(\mathrm{min})=\mathrm{Tr}[{\varrho}P({\cal H}_{sym})].
\end{equation}
This expression can be simplified if the density function
factorises, i.e. if it is of the form
\begin{equation}
q({\psi}_{1},{\ldots},{\psi}_{N})=\prod\limits_{j=1}^{N}q_{j}({\psi}_{j}).
\end{equation}
When this is the case, we can define the single-particle density
operators
\begin{equation}
{\rho}_{j}={\int}d{\sigma}({\psi})q_{j}({\psi})|{\psi}{\rangle}{\langle}{\psi}|,
\end{equation}
so that ${\varrho}={\otimes}_{j=1}^{N}{\rho}_{j}$.  This enables
us to write the minimum average inconclusive result probability as
\begin{equation}
\overline{P_{?}({\Psi})}(\mathrm{min})=\mathrm{Tr}\left[\left({\otimes}_{j=1}^{N}{\rho}_{j}\right)P({\cal H}_{sym})\right].
\end{equation}

Let us now simplify matters further by assuming that the
single-particle density functions $q_{j}({\psi})$ are all equal to
some density function $q({\psi})$.  When this is the case, the
single-particle density operators ${\rho}_{j}$ are all equal to
some density operator
${\rho}={\int}d{\sigma}({\psi})q({\psi})|{\psi}{\rangle}{\langle}{\psi}|$.
Under such circumstances it can be shown that
$\overline{P_{?}({\Psi})}(\mathrm{min})$ is the $N$th complete
symmetric polynomial in the eigenvalues of ${\rho}$ \cite{Bhatia}.
This takes an even simpler form when ${\rho}$ is the
maximally-mixed state $1_{\cal H}/D({\cal H})$, where $1_{\cal H}$
is the identity operator on ${\cal H}$. This situation arises when
the product states $|{\Psi}{\rangle}$ are uniformly-distributed,
and so the states of the $N$ particles are arbitrary and
equally-probable.  When this is the case, we obtain
\begin{equation}
\overline{P_{?}({\Psi})}(\mathrm{min})=\frac{1}{D({\cal H})^{N}}\mathrm{Tr}[P({\cal H}_{sym})]=\frac{D({\cal H}_{sym})}{D({\cal
H})^{N}}.
\end{equation}
\section{Unambiguous confirmation of all particles having different states}
\renewcommand{\theequation}{5.\arabic{equation}}
\setcounter{equation}{0} \label{sec:5}
\subsection{Requirement of linear independence}
In the preceding section, we addressed the problem of unambiguous
detection of at least one difference among the states of $N$
particles. An equally important issue to consider is the
unambiguous determination of whether or not the states of all $N$
particles are different. It is to this matter that we turn our
attentions here.

The situation we will consider here is as follows.  Each of the
$N$ particles is prepared in an element of the set of $M$ possible
pure states $\{|{\psi}_{\mu}{\rangle}\}$, like we had in section
\ref{sec:3}. Denoting the state of particle $j$ by
$|{\psi}_{{\mu}_{j}}{\rangle}$, it follows that the state of the
$N$-particle system is
$|{\Psi}_{{\mu}_{1}{\ldots}{\mu}_{N}}{\rangle}$ given by Eq.
(3.1). All of these states are taken to occur with non-zero
probability. Our aim is to devise a measurement which enables us
to unambiguously determine when all $N$ particles have been
prepared in different states.  It must give a conclusive result
with non-zero probability for the state
$|{\Psi}_{{\mu}_{1}{\ldots}{\mu}_{N}}{\rangle}$ if and only if the
${\mu}_{j}$ all have different values.

Prior to discussing this matter in detail, we should point out
that we require $M{\geq}N$.  Were this not the case, then the
number of particles would exceed the number of possible
single-particle states and so, unavoidably, at least two particles
would always be in the same state and so the states of the $N$
particles could not all be different.  It is worth making the
observation that this is a consequence of the well known
`pigeonhole principle', with particles playing the role of pigeons
and states playing that of holes.

As was the case in the preceding section, the measurement we
require to perform this task will have two possible outcomes, each
with a corresponding POVM element.  It must have a POVM element
$E_{diff}$ corresponding to unambiguous confirmation that the $N$
particles are all in different states, and a POVM element $E_{?}$
giving inconclusive results. These two operators must be positive
and satisfy the resolution of the identity in Eq. (2.3). The
probabilities of obtaining these two outcomes for a particular
$N$-particle state $|{\Psi}_{{\mu}_{1}{\ldots}{\mu}_{N}}{\rangle}$
are
\begin{eqnarray}
P_{diff}({\Psi}_{{\mu}_{1}{\ldots}{\mu}_{N}})&=&{\langle}{\Psi}_{{\mu}_{1}{\ldots}{\mu}_{N}}|E_{diff}|{\Psi}_{{\mu}_{1}{\ldots}{\mu}_{N}}{\rangle},
\\
P_{?}({\Psi}_{{\mu}_{1}{\ldots}{\mu}_{N}})&=&{\langle}{\Psi}_{{\mu}_{1}{\ldots}{\mu}_{N}}|E_{?}|{\Psi}_{{\mu}_{1}{\ldots}{\mu}_{N}}{\rangle}.
\end{eqnarray}
The result of our measurement must be unambiguous.  This means
that a conclusive result can only be obtained with non-zero
probability if the $|{\psi}_{{\mu}_{r}}{\rangle}$ or,
equivalently, the ${\mu}_{r}$, are all different.  We then require
\begin{equation}
{\langle}{\Psi}_{{\mu}_{1}{\ldots}{{\mu}_{N}}}|E_{diff}|{\Psi}_{{\mu}_{1}{\ldots}{{\mu}_{N}}}{\rangle}=0
\end{equation}
if any of the ${\mu}_{j}$ are equal.  We also require that if
$|{\psi}_{{\mu}_{1}}{\rangle},{\ldots},|{\psi}_{{\mu}_{N}}{\rangle}$
are all different then
$P_{diff}({\Psi}_{{\mu}_{1}{\ldots}{\mu}_{N}})>0$.

Here, we will show that, for a measurement satisfying the above
conditions to be possible, it is necessary that no $N$-element
subset of the set of $M$ possible single-particle states
$\{|{\psi}_{\mu}{\rangle}\}$ is linearly dependent.  Except when
$M=N$, this condition is weaker than the linear independence of
all $M$ possible single-particle states, which was our necessary
and sufficient condition for unambiguous detection of at least one
difference in section \ref{sec:3}. We will subsequently show that
the linear independence of each $N$-element subset is also a
sufficient condition.  In fact, we will show that there exists a
single, universal measurement which can be used to confirm, with
non-zero probability, that all $N$ particles have been prepared in
different states whenever this linear independence condition is
satisfied.

To prove our first assertion, let us assume that $N$ of these
states,
$|{\psi}_{{\mu}_{1}}{\rangle},{\ldots},|{\psi}_{{\mu}_{N}}{\rangle}$,
are all different, but linearly dependent.  From these states, we
construct the $N$-particle state
$|{\Psi}_{{\mu}_{1}{\ldots}{\mu}_{N}}{\rangle}$ according to Eq.
(3.1).   The linear dependence of the single-particle states
implies that the state of one of the particles may be expressed as
a linear combination of the states of the other $N-1$ particles.
We can label the $N$ particles in such a way that this particle is
particle 1.  We may then write
\begin{equation}
|{\psi}_{{\mu}_{1}}{\rangle}=\sum_{r=2}^{N}f_{r}|{\psi}_{{\mu}_{r}}{\rangle}
\end{equation}
for some complex coefficients $f_{r}$.  If we substitute this
expression for $|{\psi}_{{\mu}_{1}}{\rangle}$ into Eq. (5.1), we
obtain
\begin{equation}
P_{diff}({\Psi}_{{\mu}_{1}{\ldots}{\mu}_{N}})=\sum_{r,r'=2}^{N}f^{*}_{r'}f_{r}{\langle}{\Psi}_{{\mu}_{r'}{\mu}_{2}{\ldots}{\mu}_{N}}|E_{diff}|{\Psi}_{{\mu}_{r}{\mu}_{2}{\ldots}{\mu}_{N}}{\rangle}.
\end{equation}
We will now see that the value of this expression must be zero, by
proving that the inner products in the summation vanish.  By the
Cauchy-Schwarz inequality,
\begin{eqnarray}
&&|{\langle}{\Psi}_{{\mu}_{r'}{\mu}_{2}{\ldots}{\mu}_{N}}|E_{diff}|{\Psi}_{{\mu}_{r}{\mu}_{2}{\ldots}{\mu}_{N}}{\rangle}|^{2}
\nonumber \\
&{\leq}&{\langle}{\Psi}_{{\mu}_{r'}{\mu}_{2}{\ldots}{\mu}_{N}}|E_{diff}|{\Psi}_{{\mu}_{r'}{\mu}_{2}{\ldots}{\mu}_{N}}{\rangle}{\langle}{\Psi}_{{\mu}_{r}{\mu}_{2}{\ldots}{\mu}_{N}}|E_{diff}|{\Psi}_{{\mu}_{r}{\mu}_{2}{\ldots}{\mu}_{N}}{\rangle}
\nonumber \\
&=&P_{diff}({\Psi}_{{\mu}_{r'}{\mu}_{2}{\ldots}{\mu}_{N}})P_{diff}({\Psi}_{{\mu}_{r}{\mu}_{2}{\ldots}{\mu}_{N}}).
\end{eqnarray}
However, both factors in the last line of this expression must be
zero.  This is because they are both probabilities of obtaining a
conclusive confirmation of all particles being in different
states, but where the state of the first particle is identical to
the state of one of the other $N-1$ particles.  In the first
probability, it is particle $r'$ while in the second, it is
particle $r$.  It follows from the requirement of unambiguity that
these vanish, and so, by the Cauchy-Schwarz inequality, do the
inner products in the sum in Eq. (5.5), and therefore the sum
itself. So, we have shown that to confirm unambiguously that all
$N$ particles have been prepared in different states, it is
necessary for the states of the $N$ particles to form a linearly
independent set.  Since this must be true for every $N$-particle
initial state $|{\Psi}_{{\mu}_{1}{\ldots}{\mu}_{N}}{\rangle}$, we
see that this is equivalent to the requirement that no $N$-element
subset of the set of $M$ possible single-particle states
$\{|{\psi}_{\mu}{\rangle}\}$ is linearly dependent.

It will be shown below that this condition is also sufficient for
unambiguous confirmation that all $N$ particles are in different
states.  We stated above that this condition is, except when
$M=N$, weaker than the requirement of the set of all $M$
single-particle states to be linearly independent.  A simple
example of a set of single-particle states which satisfies the
first of these linear independence conditions but not the second
is as follows.  Suppose that $M=N+1$ and that the first $N$ of the
states $|{\psi}_{\mu}{\rangle}$ form an orthonormal basis for
${\cal H}$, but where
$|{\psi}_{M}{\rangle}=\sum_{{\mu}=1}^{N}a_{\mu}|{\psi}_{\mu}{\rangle}$,
where the $a_{\mu}$ are all non-zero.  Clearly, the set of all $M$
single-particle states $|{\psi}_{\mu}{\rangle}$ is linearly
dependent.  However, this set has no $N$-element linearly
dependent subset.  So, when the individual particles are prepared
in states in this set according to the conditions described here
and in section \ref{sec:3}, it is impossible to unambiguously
confirm that they have all been prepared in the same state,
although it is possible to unambiguously confirm that their states
are all different.
\subsection{Universal measurement strategy}
We will now show that this condition of linear independence is
also sufficient.  To do so, we will construct a measurement which,
with non-zero probability, unambiguously confirms that all $N$
particles have been prepared in different states whenever the
above linear independence condition is satisfied. For the sake of
maximum generality, we no longer restrict ourselves to the case of
each particle being prepared in one of a finite number of possible
states. Instead, we allow it to be any pure state in ${\cal H}$.
So, employing again the notation of section  \ref{sec:4}, let us
denote the initial state of particle $j$ by $|{\psi}_{j}{\rangle}$
and the entire $N$-particle state by $|{\Psi}{\rangle}$ given by
Eq. (4.1).  In the measurement we shall consider, it will be
necessary to decompose the total Hilbert space ${\cal H}_{tot}$
into two subspaces, the antisymmetric subspace, which we shall
denote by ${\cal H}_{anti}$, and its orthogonal complement, the
non-antisymmetric subspace, which we shall denote by ${\cal
H}_{na}$.  Let $P({\cal H}_{anti})$ and $P({\cal H}_{na})$ be the
projectors onto ${\cal H}_{anti}$ and ${\cal H}_{na}$
respectively.  Consider the POVM with the following elements:
\begin{eqnarray}
E_{diff}&=&P({\cal H}_{anti}), \\ E_{?}&=&P({\cal H}_{na}).
\end{eqnarray}
We will now show that this measurement satisfies our requirements.  To do so, we will derive an explicit expression for the difference
detection probability, which will be discussed in more detail in section \ref{sec:6}.  Consider again $S(N)$, the symmetric group of
degree $N$. The antisymmetric tensor product of $|{\psi}_{1}{\rangle},{\ldots},|{\psi}_{N}{\rangle}$ is defined as \cite{Bhatia}
\begin{equation}
|{\psi}_{1}{\rangle}{\wedge}{\dots}{\wedge}|{\psi}_{N}{\rangle}=\frac{1}{\sqrt{N!}}\sum_{{\sigma}{\in}S(N)}{\varepsilon}_{\sigma}|{\psi}_{{\sigma}(1)}{\rangle}{\otimes}{\dots}{\otimes}|{\psi}_{{\sigma}(N)}{\rangle},
\end{equation}
where ${\varepsilon}_{\sigma}$ is equal to $+1$ or $-1$ when
${\sigma}$ is an even or odd permutation respectively.  For the
state $|{\Psi}{\rangle}$ given by Eq. (4.1), the projector onto
the antisymmetric subspace acts as follows:
\begin{eqnarray}
P({\cal
H}_{anti})|{\Psi}{\rangle}&=&\frac{1}{\sqrt{N!}}|{\psi}_{1}{\rangle}{\wedge}{\dots}{\wedge}|{\psi}_{N}{\rangle}
\nonumber
\\ &=&\frac{1}{N
!}\sum_{{\sigma}{\in}S(N)}{\varepsilon}_{\sigma}|{\psi}_{{\sigma}(1)}{\rangle}{\otimes}{\dots}{\otimes}|{\psi}_{{\sigma}(N)}{\rangle}.
\end{eqnarray}
From this and Eqs. (2.2) and (5.8), we see that the probability of
obtaining a conclusive difference detection for the $N$-particle
state $|{\Psi}{\rangle}$ is given by
\begin{eqnarray}
P_{diff}({\Psi})&=&{\langle}{\Psi}|P({\cal H}_{anti})|{\Psi}{\rangle} \nonumber \\
&=&\frac{1}{N!}{\langle}{\psi}_{1}|{\otimes}{\dots}{\otimes}{\langle}{\psi}_{N}|\sum_{{\sigma}{\in}S(N)}{\varepsilon}_{\sigma}|{\psi}_{{\sigma}(1)}{\rangle}{\otimes}{\dots}{\otimes}|{\psi}_{{\sigma}(N)}{\rangle}
\nonumber
\\&=&\frac{1}{N!}\sum_{{\sigma}{\in}S(N)}{\varepsilon}_{\sigma}{\langle}{\psi}_{1}|{\psi}_{{\sigma}(1)}{\rangle}{\dots}{\langle}{\psi}_{N}|{\psi}_{{\sigma}(N)}{\rangle}.
\end{eqnarray}
Let us now recall the Gram matrix ${\Gamma}=({\gamma}_{ij})$ where
${\gamma}_{ij}={\langle}{\psi}_{i}|{\psi}_{j}{\rangle}$.  It
follows from Eq. (5.11) that
\begin{equation}
P_{diff}({\Psi})=\frac{1}{N!}\sum_{{\sigma}{\in}S(N)}{\varepsilon}_{\sigma}{\gamma}_{1{\sigma}(1)}{\dots}{\gamma}_{N{\sigma}(N)}.
\end{equation}
If we now make use of the following definition of the determinant
of an $N{\times}N$ matrix $A=(a_{ij})$,
\begin{equation}
\mathrm{det}(A)=\sum_{{\sigma}{\in}S(N)}{\varepsilon}_{\sigma}a_{1{\sigma}(1)}{\dots}a_{N{\sigma}(N)},
\end{equation}
we finally obtain
\begin{equation}
P_{diff}({\Psi})=\frac{1}{N!}\mathrm{det}({\Gamma}).
\end{equation}
So, for this measurement, the probability of obtaining a
conclusive result is proportional to the determinant of the Gram
matrix of the states of the $N$ particles.

Now, the determinant of a Gram matrix is non-zero if and only if
the vectors forming it are linearly independent \cite{Kreyzig}.
This proves the sufficiency of the linear independence of each
$N$-element subset of the $M$ possible single-particle states to
imply that the conclusive outcome will have a non-zero
probability. The fact that the Gram matrix determinant is zero
when the states of the $N$ particles are linearly dependent is
also sufficient to guarantee that the measurement result is
unambiguous. This is a simple consequence of the fact that,
whenever two or more particles are in the same state, the states
of the $N$ particles are clearly linearly dependent.
\subsection{Proof of optimality}
We saw above that to confirm unambiguously that all $N$ particles
have been prepared in different states, a necessary and sufficient
condition is that the states of the $N$ particles are linearly
independent.  The sufficiency part of our proof used an explicit
measurement strategy which can always achieve this confirmation
with non-zero probability if this linear independence condition is
satisfied.  This measurement is a projective measurement which
perfectly discriminates between the antisymmetric and
non-antisymmetric subspaces of ${\cal H}_{tot}$.

We will now address the issue of the optimum measurement for
detecting when all $N$ particles are in different states.  That
is, we would like to determine the measurement which attains the
minimum average probability of an inconclusive result.  As was the
case in section \ref{sec:4}, to calculate this probability, we
require knowledge of the density function
$q({\psi}_{1},{\ldots},{\psi}_{N})$. With this, we calculate the
$N$-particle density operator ${\varrho}$ using Eq. (4.13).  We
then obtain the average probability of an inconclusive result
\begin{equation}
\overline{P_{?}({\Psi})}=\mathrm{Tr}({\varrho}E_{?}),
\end{equation}
as before.

Here, we will consider the situation where the density function is
everywhere non-zero, i.e, where
$q({\psi}_{1},{\ldots},{\psi}_{N})>0$ for all
$|{\psi}_{j}{\rangle}{\in}{\cal H}$.  When this is the case, all
$N$-particle product states are possible.  We will show that,
under these circumstances, one measurement which attains the
minimum value of $\overline{P_{?}({\Psi})}$ is the measurement
described by the POVM elements in Eqs. (5.7) and (5.8).  We shall,
in fact, restrict our attention to cases when $N{\geq}3$.  The
reason for this is that, if $N=2$, then unambiguous confirmation
that both particles are in different states is equivalent to
unambiguous confirmation that they are not both in the same state.
This problem was solved in the preceding section, and it was found
that the optimum measurement is a projective measurement with the
POVM elements $E_{?}=P({\cal H}_{sym})$ and $E_{diff}=P({\cal
H}_{asym})$. However, for $N=2$, we have ${\cal H}_{sym}={\cal
H}_{na}$ and ${\cal H}_{asym}={\cal H}_{anti}$, so for the special
case of $N=2$ we have already proven the above claim (indeed under
more restrictive conditions) in section \ref{sec:4}.

In proving this claim for $N{\geq}3$,  we shall first prove that under the specified conditions, the requirement of unambiguity implies
that the support of $E_{diff}$ must be a subspace of ${\cal H}_{anti}$. We will then show that, under conditions of optimality,
$E_{diff}$ may be taken to be the projector onto ${\cal H}_{anti}$.

To prove the first of these points, let $|e{\rangle}$ be an
eigenstate of $E_{diff}$ with non-zero eigenvalue. For any
$N$-particle product state $|{\Psi}{\rangle}$ where at least two
of the $|{\psi}_{j}{\rangle}$ are identical, the positivity of
$E_{diff}$ and the requirement of unambiguity imply that we must
have
\begin{equation}
{\langle}{\Psi}|e{\rangle}=0.
\end{equation}
To proceed, let us partition the set of $N$ particles into two subsets, ${\alpha}$ and $\bar{\alpha}$.  The former consists of any pair
of particles and the latter consists of the remaining $N-2$.  Let us denote the Hilbert spaces of these multiparticle systems by ${\cal
H}_{\alpha}$ and ${\cal H}_{\bar\alpha}$ respectively.  We can express any $N$-particle pure state in Schmidt decomposition form, where
each element of the Schmidt basis is a tensor product of one state from each of the spaces ${\cal H}_{\alpha}$ and ${\cal
H}_{\bar\alpha}$. Applying this idea to the state $|e{\rangle}$, we see that we may write
\begin{equation}
|e{\rangle}=\sum_{r=1}^{D^{2}}c_{r}|x_{r}{\rangle}{\otimes}|y_{r}{\rangle}.
\end{equation}
Here, the set $\{|x_{r}{\rangle}\}$ is an orthonormal basis for
${\cal H}_{\alpha}$ and the $|y_{r}{\rangle}$ are $D^{2}$
orthonormal states in ${\cal H}_{\bar\alpha}$.  Now suppose that
the states of the two particles in ${\alpha}$ are identical.  If
this is so, then we can write
$|{\Psi}{\rangle}=|{\psi}{\rangle}{\otimes}|{\psi}{\rangle}{\otimes}|{\Phi}{\rangle}$
for some $|{\psi}{\rangle}{\otimes}|{\psi}{\rangle}{\in}{\cal
H}_{\alpha}$ and some  $(N-2)$-particle product state
$|{\Phi}{\rangle}{\in}{\cal H}_{\bar\alpha}$.  Substituting this
into Eq. (5.16) and making use of Eq. (5.17) we obtain
\begin{equation}
\sum_{r=1}^{D^{2}}c_{r}({\langle}{\psi}|{\otimes}{\langle}{\psi}|)|x_{r}{\rangle}{\langle}{\Phi}|y_{r}{\rangle}=0.
\end{equation}
This must be true for all $(N-2)$-particle product states
$|{\Phi}{\rangle}{\in}{\cal H}_{\bar\alpha}$ and for all
$|{\psi}{\rangle}{\in}{\cal H}$. From Eq. (5.18), we can deduce
that for any $r$ such that $c_{r}{\neq}0$, we have
$({\langle}{\psi}|{\otimes}{\langle}{\psi}|)|x_{r}{\rangle}=0$ for
all $|{\psi}{\rangle}{\in}{\cal H}$.  It is quite simple to prove
this.  Consider the vector
$\sum_{r'}c_{r'}[({\langle}{\psi}|{\otimes}{\langle}{\psi}|)|x_{r'}{\rangle}]|y_{r'}{\rangle}{\in}{\cal
H}_{\bar\alpha}$.  From Eq. (5.18), we see that this vector is
orthogonal to all product states in ${\cal H}_{\bar{\alpha}}$.
Since any vector in ${\cal H}_{\bar{\alpha}}$ can be expressed as
a linear combination of product states in ${\cal
H}_{\bar{\alpha}}$, it must therefore be orthogonal to all vectors
in ${\cal H}_{\bar{\alpha}}$.  This can only be true if it is the
zero vector:
\begin{equation}
\sum_{r'=1}^{D^{2}}c_{r'}[({\langle}{\psi}|{\otimes}{\langle}{\psi}|)|x_{r'}{\rangle}]|y_{r'}{\rangle}=0.
\end{equation}
If we now take the inner product throughout this expression with
any $|y_{r}{\rangle}$ for which $c_{r}{\neq}0$, we finally obtain
$({\langle}{\psi}|{\otimes}{\langle}{\psi}|)|x_{r}{\rangle}=0$ for
all $|{\psi}{\rangle}{\in}{\cal H}$, as we claimed.

From Eq. (4.15), we see that the projector onto the symmetric
subspace of ${\cal H}_{\alpha}$ is ${D({\cal H})+1 \choose
2}{\int}d{\sigma}({\psi})(|{\psi}{\rangle}{\langle}{\psi}|)^{{\otimes}2}$.
Combining this expression with
$({\langle}{\psi}|{\otimes}{\langle}{\psi}|)|x_{r}{\rangle}=0$, it
follows
 that if $c_{r}{\neq}0$ then $|x_{r}{\rangle}$ is orthogonal to the symmetric subspace of ${\cal H}_{\alpha}$. It must therefore be an
element of its orthogonal complement. For a bipartite quantum system such as ${\alpha}$, the orthogonal complement of the symmetric
subspace is the antisymmetric subspace.  This implies that the state $|e{\rangle}$ is antisymmetric under exchange of the two particles
in ${\alpha}$. However, the set ${\alpha}$ comprises an arbitrary pair chosen from all $N$ particles, and so the state $|e{\rangle}$
must be antisymmetric under exchange of any pair of particles. This implies that it must lie in the $N$-particle antisymmetric subspace
${\cal H}_{anti}$. Finally, since $|e{\rangle}$ is an arbitrary eigenstate of $E_{diff}$ with non-zero eigenvalue, it follows that all
eigenstates of this operator with non-zero eigenvalue lie in ${\cal H}_{anti}$.  In other words, the support of $E_{diff}$ is a
subspace of ${\cal H}_{anti}$.

To prove the second assertion, we make use of the fact that, in
view of the above argument, $E_{diff}$ can be spectrally
decomposed in terms of an orthonormal basis for ${\cal H}_{anti}$:
\begin{equation}
E_{diff}=\sum_{r=1}^{D({\cal H}_{anti})}e_{r}|e_{r}{\rangle}{\langle}e_{r}|.
\end{equation}
Here, $0{\leq}e_{r}{\leq}1$ and the $|e_{r}{\rangle}$ form an orthonormal basis for ${\cal H}_{anti}$.  Also, $D({\cal H}_{anti})$ is
the dimensionality of ${\cal H}_{anti}$ and it has the value
\begin{equation}
D({\cal H}_{anti})={D({\cal H}) \choose N}.
\end{equation}
Substitution of the expression in Eq. (5.20) for $E_{diff}$ into Eq. (5.15) and making use of the resolution of the identity in Eq.
(2.3) gives
\begin{eqnarray}
\overline{P_{?}({\Psi})}&=&\mathrm{Tr}({\varrho}E_{?})=1-\mathrm{Tr}({\varrho}E_{diff}) \nonumber \\
&=&1-\sum_{r=1}^{D({\cal H}_{anti})}e_{r}{\langle}e_{r}|{\varrho}|e_{r}{\rangle} \nonumber \\
&{\geq}&1-\sum_{r=1}^{D({\cal H}_{anti})}{\langle}e_{r}|{\varrho}|e_{r}{\rangle} \nonumber \\
&=&1-\mathrm{Tr}[{\varrho}P({\cal H}_{anti})].
\end{eqnarray}
So, the minimum value of $\overline{P_{?}({\Psi})}$ occurs when
the difference detection POVM element $E_{diff}$ is the projector
onto the antisymmetric subspace ${\cal H}_{anti}$. From the
resolution of the identity in Eq. (2.3), we see that the
corresponding POVM element for inconclusive results, $E_{?}$, must
be equal to $P({\cal H}_{na})$, the projector onto the
non-antisymmetric subspace ${\cal H}_{na}$. We have shown that,
under the specified conditions, the measurement described by the
POVM elements in Eqs. (5.7) and (5.8) is optimal and that
\begin{equation}
\overline{P_{?}({\Psi})}(\mathrm{min})=\mathrm{Tr}[{\varrho}P({\cal H}_{na})].
\end{equation}
As was the case for the measurement described in the preceding
section for universally detecting a single difference, this
probability can be simplified if the density function factorises
according to Eq. (4.21).  When this is so, the $N$-particle
density operator takes the form
${\varrho}={\otimes}_{j=1}^{N}{\rho}_{j}$ where the
single-particle density operators ${\rho}_{j}$ are given by Eq.
(4.22). This enables us to write the minimum average inconclusive
result probability as
\begin{equation}
\overline{P_{?}({\Psi})}(\mathrm{min})=1-\mathrm{Tr}\left[\left({\otimes}_{j=1}^{N}{\rho}_{j}\right)P({\cal H}_{anti})\right].
\end{equation}
In particular, if the  ${\rho}_{j}$ are all equal to some
single-particle density operator ${\rho}$, then the last term in
this above expression has the form
$\mathrm{Tr}[{\rho}^{{\otimes}N}P({\cal H}_{anti})]$. This can be
shown to be equal to the $N$th elementary symmetric polynomial in
the eigenvalues of ${\rho}$ \cite{Bhatia}. Finally, if ${\rho}$ is
the maximally-mixed state $1_{\cal H}/D({\cal H})$, then we obtain
\begin{equation}
\overline{P_{?}({\Psi})}(\mathrm{min})=1-\frac{1}{D({\cal H})^{N}}\mathrm{Tr}[P({\cal H}_{anti})]=1-\frac{D({\cal H}_{anti})}{D({\cal
H})^{N}}.
\end{equation}
This situation arises when the $N$-particle product states
$|{\Psi}{\rangle}$ are uniformly-distributed, in which case the
states of the $N$ particles are arbitrary and equally-probable.
\section{Optimal universal difference detection and distinguishability}
\renewcommand{\theequation}{6.\arabic{equation}}
\setcounter{equation}{0} \label{sec:6} In sections \ref{sec:4} and
\ref{sec:5}, we considered optimal, universal difference
detection. The aim in both sections was to determine measurement
strategies which optimally detect differences among the states of
$N$ particles. For a broad range of circumstances, including when
all $N$-particle product states are possible, we obtained the
general form of the optimal strategies for detecting when at least
two, and when all $N$ particles are in different states.  We also
obtained expressions for the associated difference detection
probabilities. For the measurement used to unambiguously
determinine  when the states of all $N$ particles are different,
the success probability for a particular $N$-particle product
state $|{\Psi}{\rangle}={\otimes}_{j=1}^{N}|{\psi}_{j}{\rangle}$
is given by Eq. (5.14), which we repeat here for convenience:
\begin{equation}
P_{diff}({\Psi})=\frac{1}{N!}\mathrm{det}(\Gamma).
\end{equation}
For the measurement which detects when at least two particles are
in different states, the inconclusive result probability for a
particular $N$-particle product state is given by Eq. (4.10).
From this, we see that the success probability is
\begin{equation}
P_{diff}({\Psi})=1-\frac{1}{N!}\mathrm{per}(\Gamma).
\end{equation}
For these measurements, the true figures of merit are the averages
of these quantities over all product states with respect to the
density function $q$.  These probabilities are equal to 1 minus
the corresponding minimum average inconclusive result
probabilities, given by Eqs. (5.25) and (4.24) respectively.
Nevertheless, the success probabilities in Eqs. (6.1) and (6.2)
for a particular $N$-particle state do depend on the states of the
individual particles through their Gram matrix ${\Gamma}$. In this
section, we shall examine these success probabilities in more
detail. Our focus will be on their state-dependent nature. It is
natural to enquire as to what property of the states of the
individual particles these probabilities relate to.  Here, we
shall investigate the possibility that, for a fixed $N$-particle
product state $|{\Psi}{\rangle}$, these success probabilities can
be viewed as measures of the distinguishability of the states of
the individual particles $|{\psi}_{j}{\rangle}$, considered as
possible states of a single particle. Our intuition is that, the
higher the probability of obtaining a conclusive result, the more
distinguishable these states are.  Often, in discussions of
distinguishability, the a priori probabilities of the states, as
well as the states themselves, are taken into account.  The reason
why is that distinguishability measures are typically
probabilities, which we need to know the a priori probabilities of
the possible states to evaluate.  Here, we simply take all states
to be equally-probable.

To explore the validity of this idea, it is necessary to determine
whether or not the probabilities in Eqs. (6.1) and (6.2) have the
properties which any distinguishability measure must have.  As
yet, no general set of necessary and sufficient criteria for a
given quantity to be a suitable distinguishability measure have,
to our knowledge, been proposed. However, certain intuitive
necessary conditions are widely accepted.  One is the fact that
only orthogonal sets of states can be considered
maximally-distinguishable.  Another is the fact that a
distinguishability measure must be non-increasing under a
deterministic quantum operation.  We show that both of these
success probabilities satisfy the first of these criteria. We also
find that the second criterion is always satisfied by the
probability of finding all states being different in Eq. (6.1).
Whether or not this is also true in general for the success
probability of finding at least one difference among the states in
Eq. (6.2) is, we find, equivalent to an open conjecture in linear
algebra.

We will first prove that the success probability in Eq. (6.1) for
finding all states being different satisfies the above criteria.
We will then carry out a corresponding analysis for the
probability of confirming at least one difference in Eq. (6.2).
Beginning with $P_{diff}({\Psi})$ in Eq. (6.1), we are interested
in whether or not this attains its maximum value when the states
in the Gram matrix ${\Gamma}$ are orthogonal.  This is clearly
equivalent to asking if the Gram matrix determinant maximises for
orthogonal states. This is indeed the case, and to prove it, we
use Hadamard's inequality \cite{HJ}, which is satisfied by any
positive  $N{\times}N$ matrix $A=(a_{ij})$,
\begin{equation}
\mathrm{det}(A){\leq}\prod\limits_{i=1}^{N}a_{ii},
\end{equation}
where the equality is satisfied if and only if $A$ is diagonal.

All Gram matrices are positive.  For normalised states, the
diagonal elements of a Gram matrix are all equal to 1. It follows
from the Hadamard inequality that $\mathrm{det}({\Gamma}){\leq}1$,
with the equality holding if and only if the off-diagonal elements
of ${\Gamma}$ vanish. This corresponds to the states
$|{\psi}_{j}{\rangle}$ being orthogonal and therefore
maximally-distinguishable. So, $P_{diff}({\Psi})$ in Eq. (6.1),
considered as a distinguishability measure, correctly identifies
the maximally-distinguishable sets of states as being orthogonal
sets.

We saw in section \ref{sec:5} that the determinant of ${\Gamma}$
takes the value of zero if and only if the single-particle states
are linearly dependent.  For a positive matrix such as ${\Gamma}$,
this is the minimum value of $\mathrm{det}({\Gamma})$.   However,
a linearly dependent set of states does not necessarily have zero
distinguishability for all state discrimination strategies. For
example, in quantum hypothesis testing, where we aim to
discriminate among the states with minimum error probability,
linearly dependent sets do not in general represent the worst case
scenario.  However, they do in other state discrimination
strategies. In particular, in unambiguous state discrimination,
the probability of discriminating among a linearly dependent set
of states is zero.  However, this is not particularly important.
It is known that different distinguishability measures do not, in
general, characterise the distinguishability of the same set of
states in the same way.  In particular, it is known that different
distinguishability measures can impose different orderings on two
sets of states.  This matter is explored in more detail in
\cite{Chefles3}.

What this implies is that, if we wish to further test the Gram
matrix determinant for its viability as a distinguishability
measure, then we should be judicious about which aspects of its
behaviour we should assess. Quantum state discrimination has an
operational aspect which is, to a certain extent, distinct from
the actual act of distinguishing among quantum states.  This is a
constraint on the transformation of a distinguishability measure
under a deterministic quantum operation. All distinguishability
measures must be non-increasing under such operations.  The reason
for this is simple.   If this condition were not satisfied, then
we could increase the value of the distinguishability measure by
incorporation of a deterministic quantum operation that increases
it into the state discrimination procedure.  Deterministic quantum
operations are represented mathematically by completely positive,
linear, trace-preserving (CPLTP) maps. Quantities which have this
non-increasing property are known as distinguishability monotones
\cite{Chefles4}.

Clearly, it is of interest to determine whether or not the Gram matrix determinant is a distinguishability monotone.  For this question
to be meaningful, we must restrict our attention to situations where the initial pure states, which we shall write as
$|{\psi}^{1}_{j}{\rangle}$, are transformed into another set of pure states, $|{\psi}^{2}_{j}{\rangle}$, since the Gram matrix is
defined only for pure states.

It is known \cite{Uhlmann,Chefles2,Chefles4,CJW} that a necessary and sufficient condition for the existence of a CPLTP map which
deterministically transforms $|{\psi}^{1}_{j}{\rangle}$ into $|{\psi}^{2}_{j}{\rangle}$ for each $j=1,{\ldots},N$ is that there exists
an $N{\times}N$ positive matrix ${\Pi}$ such that
\begin{equation}
{\Gamma}_{1}={\Gamma}_{2}{\circ}{\Pi},
\end{equation}
where ${\Gamma}_{1}$ and ${\Gamma}_{2}$ are the Gram matrices of the initial and final sets of states respectively and `${\circ}$'
denotes the Hadamard product.  The Hadamard product $A{\circ}B$ of two $N{\times}N$ matrices  $A=(a_{ij})$ and $B=(b_{ij})$ has $ij$
element $a_{ij}b_{ij}$, i.e., it is the entrywise product. We would like to know if Eq. (6.4) implies that
\begin{equation}
\mathrm{det}({\Gamma}_{1}){\geq}\mathrm{det}({\Gamma}_{2}),
\end{equation}
that is, whether or not the Gram matrix determinant is
non-increasing under such a transformation.  We shall now prove
that this inequality is indeed always satisfied for such a
transformation.  To do so, we shall use Oppenheim's inequality
\cite{HJ,Oppenheim}. For any two positive $N{\times}N$ matrices
$A=(a_{ij})$ and $B=(b_{ij})$, Oppenheim proved that
\begin{equation}
\mathrm{det}(A{\circ}B){\geq}\mathrm{det}(A)\prod\limits_{i=1}^{N}b_{ii}{\geq}\mathrm{det}(A)\mathrm{det}(B).
\end{equation}
Notice that the second inequality here follows as a consequence of the Hadamard inequality in (6.3).   To make use of this, let
$A={\Gamma}_{2}$ and $B={\Pi}$. If Eq. (6.4) is satisfied, then $A{\circ}B={\Gamma}_{1}$.  From this and the first inequality in (6.6)
we see that
\begin{equation}
\mathrm{det}({\Gamma}_{1})=\mathrm{det}({\Gamma}_{2}{\circ}{\Pi}){\geq}\mathrm{det}({\Gamma}_{2})\prod\limits_{i=1}^{N}{\pi}_{ii},
\end{equation}
where ${\pi}_{ii}$ is the $ii$ element of ${\Pi}$.  The diagonal
elements of the Gram matrices ${\Gamma}_{1}$ and ${\Gamma}_{2}$
are all equal to 1. From this and Eq. (6.4), it follows that the
${\pi}_{ii}$ must also all be equal to 1, as must be their
product.  We therefore obtain inequality (6.5) and this completes
the proof.

An elementary, but illustrative example of this theorem is the
simple case of $N=2$.  Let the $|{\psi}^{1}_{j}{\rangle}$ be the
initial pair of states and $|{\psi}^{2}_{j}{\rangle}$ be the final
pair.  Then
\begin{equation}
\mathrm{det}({\Gamma}_{r})=1-|{\langle}{\psi}_{1}^{r}|{\psi}_{2}^{r}{\rangle}|^{2}.
\end{equation}
Combining this with inequality (6.5), we see that it is equivalent to the overlap of $|{\psi}_{1}^{1}{\rangle}$ and
$|{\psi}_{2}^{1}{\rangle}$ being no greater than the overlap of $|{\psi}_{1}^{2}{\rangle}$ and $|{\psi}_{2}^{2}{\rangle}$.  So, in the
simple case of $N=2$, inequality (6.5) expresses the non-decreasing nature of the overlap of the pair of states.

Let us now carry out a corresponding analysis for the difference
detection probability in Eq. (6.2).  Here, $P_{diff}({\Psi})$ is
the probability of successfully confirming that the states of all
$N$ particles are different.  Could this be a suitable measure of
the distinguishability of the single-particle states?

This question is, of course, equivalent to asking if the
corresponding minimum probability of obtaining an inconclusive
result for this measurement, $P_{?}({\Psi})$ in Eq. (4.10), and
therefore the Gram matrix permanent, is a suitable measure of the
indistinguishability of the single-particle states. For reasons
that will become apparent, it will turn out to be more convenient
to investigate the matter from this perspective. If the Gram
matrix permanent is to be a suitable measure of the
indistinguishability of a set of states, then intuitively it
should take its minimum value when the states are orthogonal. To
establish this, it would be helpful if there existed an inequality
for the permanent which could play a role analogous to that played
by Hadamard's inequality in our discussion of the determinant.
Such an inequality exists and it is known as Marcus's inequality
\cite{Marcus}.  For any positive $N{\times}N$ matrix $A=(a_{ij})$,
the following inequality holds:
\begin{equation}
\mathrm{per}(A){\geq}\prod\limits_{i=1}^{N}a_{ii}.
\end{equation}
This is the permanental analogue of the Hadamard inequality (6.3) and, again, the equality is satisfied if and only $A$ is diagonal. If
$A={\Gamma}$, then we have $\mathrm{per}({\Gamma}){\geq}1$, where the equality is attained only in the case where the off-diagonal
elements of ${\Gamma}$ vanish, i.e. when the single-particle states are orthogonal.  This confirms that $\mathrm{per}({\Gamma})$, as an
indistinguishability measure, correctly identifies orthogonal states as the least indistinguishable, or equivalently, most
distinguishable states. Furthermore, as we saw in section \ref{sec:4}, $\mathrm{per}({\Gamma})$ takes its maximum value of $N!$ if and
only if the states are all identical.  So, it correctly identifies a set of identical states as having minimum (zero)
distinguishability. We saw in our discussion of the Gram matrix determinant that, for a particular state discrimination strategy, it is
not, in general, necessary for a set of states to be identical to have zero distinguishability.  However, this condition is unarguably
sufficient.

If the Gram matrix permanent is to be a suitable
indistinguishability measure, then, under deterministic
transformations of one set of pure states into another, it must
display the opposite kind of behaviour to the Gram matrix
determinant, i.e. it must be non-decreasing under any such
transformation. That is, whenever Eq. (6.4) holds, we require that
\begin{equation}
\mathrm{per}({\Gamma}_{1}){\leq}\mathrm{per}({\Gamma}_{2}).
\end{equation}
It could be proven that the Gram matrix permanent is non-decreasing under such transformations if there existed a suitable permanental
analogue of Oppenheim's inequality. Unfortunately, whether or not the required inequality is true in general is, at the time of
writing, an open question.

The mathematical problem  of finding a permanental analogue of
Oppenheim's inequality has an interesting history.  The first step
in this direction was made by Chollet \cite{Chollet}, who
conjectured that for any two $N{\times}N$ positive  matrices $A$
and $B$,
\begin{equation}
\mathrm{per}(A{\circ}B){\leq}\mathrm{per}(A)\mathrm{per}(B).
\end{equation}
Subsequently, Bapat and Sunder \cite{Bapat1} made a stronger
conjecture than Chollet.  They proposed that for any two positive
 $N{\times}N$ matrices $A$ and $B$,
\begin{equation}
\mathrm{per}(A{\circ}B){\leq}\mathrm{per}(A)\prod\limits_{i=1}^{N}b_{ii}{\leq}\mathrm{per}(A)\mathrm{per}(B),
\end{equation}
where the second inequality here is easily seen to hold as a consequence of Marcus's inequality (6.9). As a consequence of this, the
Bapat-Sunder conjecture is equivalent to
\begin{equation}\mathrm{per}(A{\circ}B){\leq}\mathrm{per}(A)\prod\limits_{i=1}^{N}b_{ii}.
\end{equation}
Bapat and Sunder have subsequently shown that this inequality is
satisfied by all $2{\times}2$ and $3{\times}3$ positive
 matrices \cite{Bapat2}.  The weaker conjecture (6.11) of Chollet was also proven for such matrices independently by Gregorac and
Hentzel \cite{Gregorac}.  Furthermore, Bapat and Sunder \cite{Bapat2} and Grone and Merris \cite{Grone} have discovered interesting
connections between the Bapat-Sunder conjecture and other conjectured identities relating to permanents.

Here, we will show that if the Bapat-Sunder conjecture is indeed
true for all $N{\times}N$ positive  matrices $A$ and $B$, then
inequality (6.10) always holds. Assuming the general validity of
this conjecture, let us make the same substitutions as before,
that is, $A={\Gamma}_{2}$ and $B={\Pi}$. If Eq. (6.4) is true,
then
\begin{equation}
\mathrm{per}({\Gamma}_{1})=\mathrm{per}({\Gamma}_{2}{\circ}{\Pi}){\leq}\mathrm{per}({\Gamma}_{2})\prod\limits_{i=1}^{N}{\pi}_{ii},
\end{equation}
where ${\pi}_{ii}$ is the $ii$ element of ${\Pi}$.  The diagonal
elements of the Gram matrices are all equal to 1.  From Eq. (6.4),
it follows that the ${\pi}_{ii}$ must also all be equal to 1, as
must be their product, which leads to (6.10).

For the simple case of $N=2$, we have
\begin{equation}
\mathrm{per}({\Gamma}_{r})=1+|{\langle}{\psi}_{1}^{r}|{\psi}_{2}^{r}{\rangle}|^{2}.
\end{equation}
Comparing this with Eq. (6.8), we see that the non-decreasing nature of the permanent of the Gram matrix is, like the non-increasing
nature of its determinant, equivalent to the non-decreasing of the overlap of a pair of pure states.

For $N>3$ states, it is clear that this argument rests on whether
or not the Bapat-Sunder conjecture is true.  The fact that these
authors proved their conjecture for all $N{\times}N$ positive
matrices where $N{\leq}3$ implies that the Gram matrix permanent
is indeed non-decreasing under pure${\rightarrow}$pure state set
transformations for $N{\leq}3$.  Furthermore, it can be shown that
the validity of (6.10) for every pair of $N{\times}N$ Gram
matrices satisfying Eq. (6.4) is actually equivalent to the
ostensibly more general Bapat-Sunder conjecture \cite{BapatPC}. In
other words, the question of whether or not the probability of a
conclusive result for the universal strategy for finding at at
least two states to be different is a distinguishability monotone
is fully equivalent to a conjectured property of $N{\times}N$
matrices which has been resolved affirmatively for $N{\leq}3$ but
is open for $N>3$.

One might think that (6.10) could be established using the properties of general CPLTP maps, such as, for example, their contractivity,
rather than focusing on the particular properties of  transformations between sets of pure states.  It would be possible to prove
(6.10) if, under a general CPLTP map ${\cal E}$ applied separately to $N$ systems prepared in an arbitrary initial product state, the
expectation value of the symmetric subspace projector was non-decreasing, that is
\begin{equation}
\mathrm{Tr}\left[\left({\cal E}({\rho}_{1}){\otimes}{\ldots}{\otimes}{\cal E}({\rho}_{N})\right)P({\cal
H}_{sym})\right]{\geq}\mathrm{Tr}\left[\left({\rho}_{1}{\otimes}{\ldots}{\otimes}{\rho}_{N}\right)P({\cal H}_{sym})\right],
\end{equation}
where the ${\rho}_{j}=|{\psi}_{j}{\rangle}{\langle}{\psi}_{j}|$
are arbitrary pure states in ${\cal H}$. However, this is not the
case. As a counterexample, let ${\cal E}$ be the CPLTP map which
transforms every density operator on ${\cal H}$ into the
maximally-mixed state and let each
$|{\psi}_{j}{\rangle}=|{\psi}{\rangle}$ for some
$|{\psi}{\rangle}{\in}{\cal H}$. Then, clearly the $N$-particle
initial state lies in the symmetric subspace and so the r.h.s. of
(6.16) is equal to 1. However, the l.h.s. is equal to $D({\cal
H}_{sym})/D({\cal H})^{N}<1$ for $D({\cal H})>1$.  So, for this
operation ${\cal E}$, the expectation value of the symmetric
subspace projector can decrease, disproving the validity of (6.16)
in general.  As a consequence of this, we see that the expectation
value of the symmetric subspace projector is not a suitable
indistinguishability measure for mixed states.

\section{Unambiguous overlap filtering}
\renewcommand{\theequation}{7.\arabic{equation}}
\setcounter{equation}{0} \label{sec:7} So far in this paper, we have been discussing the issue of unambiguous state comparison, which
involves determining whether or not a number of quantum systems have been prepared in the same state. We have focused on the simple
case where all possible states of the individual systems are pure. When this is so, we can view state comparison as the determination
of whether or not the overlaps of the states of pairs of systems are equal to 1.

This suggests the following generalisation: we would like to
determine, unambiguously, whether or not the overlap of the states
of a pair of systems is equal to some ${\omega}$, where ${\omega}$
is a real parameter that can take any fixed value between 0 and 1.
We shall term this procedure unambiguous overlap filtering. The
problem of measuring overlaps, and relatedly, fidelities has
recently received some attention.  Winter \cite{Winter} has shown
that there is no operational procedure for measuring the fidelity,
or equivalently the overlap, of the states of two quantum systems.
More recently, Ekert et al \cite{Functionals} have discussed
interesting relationships between fidelity/overlap estimation and
a number of other important tasks in quantum information
processing.

Consider two particles, 1 and 2, each with a copy of the $D({\cal
H})$ dimensional Hilbert space ${\cal H}$. The Hilbert space of
the total system is ${\cal H}_{tot}={\cal H}{\otimes}{\cal H}$.
Throughout this section, the left and right tensor factors will
refer to particles 1 and 2 respectively.  Let 1 and 2 be initially
prepared in the pure states $|{\psi}_{1}{\rangle}$ and
$|{\psi}_{2}{\rangle}$ respectively. These states are taken to be
completely arbitrary and unknown pure states in ${\cal H}$.
 In unambiguous state comparison, our aim is to determine with zero probability of error but with the possibility of an inconclusive
result whether or not the states $|{\psi}_{1}{\rangle}$ and
$|{\psi}_{2}{\rangle}$ are identical.  In other words, we aim to
determine whether or not the overlap of these states,
$|{\langle}{\psi}_{1}|{\psi}_{2}{\rangle}|$, is equal to 1. In
unambiguous overlap filtering, our aim is to decide unambiguously
whether or not
$|{\langle}{\psi}_{1}|{\psi}_{2}{\rangle}|={\omega}$ for some
arbitrary fixed ${\omega}{\in}[0,1]$.

We would like a measurement whose result tells us, unambiguously,
that the overlap is, or is not equal to ${\omega}$, with some
non-zero probability.  We then require a filter which takes as its
inputs particles 1 and 2 prepared in the unknown pure states
$|{\psi}_{1}{\rangle}$ and $|{\psi}_{2}{\rangle}$ respectively.
The measurement is required to unambiguously distinguish the cases
$|{\langle}{\psi}_{1}|{\psi}_{2}{\rangle}|={\omega}$ and
$|{\langle}{\psi}_{1}|{\psi}_{2}{\rangle}|{\neq}{\omega}$

This measurement will have three possible outcomes: `yes', which
signals that the overlap is equal to ${\omega}$, `no', which
signals that the overlap is not equal to ${\omega}$, and `?',
which is an inconclusive result.  These requirements imply that,
formally, the measurement will be described by a three-element
POVM. The three elements are $E_{yes}({\omega})$,
$E_{no}({\omega})$ and $E_{?}({\omega})$. These are positive
operators on the space ${\cal H}_{tot}$, and the unambiguous
nature of their outcomes implies the conditions:
\begin{eqnarray}
{\langle}{\psi}_{1}|{\otimes}{\langle}{\psi}_{2}|E_{yes}({\omega})|{\psi}_{1}{\rangle}{\otimes}|{\psi}_{2}{\rangle}&{\neq}&0{\Rightarrow}|{\langle}{\psi}_{2}|{\psi}_{1}{\rangle}|={\omega},
\\
{\langle}{\psi}_{1}|{\otimes}{\langle}{\psi}_{2}|E_{no}({\omega})|{\psi}_{1}{\rangle}{\otimes}|{\psi}_{2}{\rangle}&{\neq}&0{\Rightarrow}|{\langle}{\psi}_{2}|{\psi}_{1}{\rangle}|{\neq}{\omega}.
\end{eqnarray}
They must also resolve the identity:
\begin{equation}
E_{yes}({\omega})+E_{no}({\omega})+E_{?}({\omega})=1_{tot}.
\end{equation}
If we obtain the result `yes', then we know for sure that
$|{\langle}{\psi}_{2}|{\psi}_{1}{\rangle}|={\omega}$. If, on the
other hand, we obtain the result `no', then we know for sure that
$|{\langle}{\psi}_{2}|{\psi}_{1}{\rangle}|{\neq}{\omega}$.

We will now prove that, if we require Eqs. (7.1) and (7.2) to hold
for all $|{\psi}_{1}{\rangle},|{\psi}_{2}{\rangle}{\in}{\cal H}$,
then
\begin{eqnarray}
E_{yes}({\omega})&=&0\;{\forall}\;{\omega}{\in}[0,1], \\
E_{no}({\omega})&{\neq}&0{\Rightarrow}{\omega}=1.
\end{eqnarray}
To prove Eq. (7.4), let ${\cal H}_{12}$ be a two dimensional
subspace of ${\cal H}$ which contains both $|{\psi}_{1}{\rangle}$
and $|{\psi}_{2}{\rangle}$.  When $|{\psi}_{1}{\rangle}$ and
$|{\psi}_{2}{\rangle}$ are not identical, ${\cal H}_{12}$ is
clearly uniquely defined as the subspace spanned by
$|{\psi}_{1}{\rangle}$ and $|{\psi}_{2}{\rangle}$.  Let
$\{|x_{1}{\rangle},|x_{2}{\rangle}\}$ and
$\{|y_{1}{\rangle},|y_{2}{\rangle}\}$ be, possibly nonorthogonal,
basis sets for  ${\cal H}_{12}$ such that
\begin{equation}
|{\langle}x_{k}|y_{l}{\rangle}|{\neq}{\omega}.
\end{equation}
Such basis sets can always be constructed.  We may write
\begin{equation}
|{\psi}_{j}{\rangle}=\sum_{k=1}^{2}a_{jk}|x_{k}{\rangle}=\sum_{l=1}^{2}b_{jl}|y_{l}{\rangle},
\end{equation}
where $j=1,2$.   Suppose now that $|{\langle}{\psi}_{2}|{\psi}_{1}{\rangle}|={\omega}$ and calculate
\begin{equation}
{\langle}{\psi}_{1}|{\otimes}{\langle}{\psi}_{2}|E_{yes}({\omega})|{\psi}_{1}{\rangle}{\otimes}|{\psi}_{2}{\rangle}=\sum_{k,k',l,l'=1}^{2}a^{*}_{1k'}b^{*}_{2l'}a_{1k}b_{2l}{\langle}x_{k'}|{\otimes}{\langle}y_{l'}|E_{yes}({\omega})|x_{k}{\rangle}{\otimes}|y_{l}{\rangle}.
\end{equation}
\\
As we shall now see, all  matrix elements of $E_{yes}({\omega})$
in this sum are vanishing.  To prove this, we use the
Cauchy-Schwarz inequality, which gives
\begin{equation}
|{\langle}x_{k'}|{\otimes}{\langle}y_{l'}|E_{yes}({\omega})|x_{k}{\rangle}{\otimes}|y_{l}{\rangle}|^{2}{\leq}{\langle}x_{k'}|{\otimes}{\langle}y_{l'}|E_{yes}({\omega})|x_{k'}{\rangle}{\otimes}|y_{l'}{\rangle}{\langle}x_{k}|{\otimes}{\langle}y_{l}|E_{yes}({\omega})|x_{k}{\rangle}{\otimes}|y_{l}{\rangle}=0,
\end{equation}
as a consequence of combining Eq. (7.1) with Eq. (7.6).  We then
conclude that
${\langle}{\psi}_{1}|{\otimes}{\langle}{\psi}_{2}|E_{yes}({\omega})|{\psi}_{1}{\rangle}{\otimes}|{\psi}_{2}{\rangle}=0$
for all $|{\psi}_{1}{\rangle},|{\psi}_{2}{\rangle}{\in}{\cal H}$.
This can be used to prove that $E_{yes}({\omega})=0$ in the
following way: we calculate the trace of $E_{yes}({\omega})$ in a
product basis and from Eqs. (7.8) and (7.9) it follows that
$\mathrm{Tr}[E_{yes}({\omega})]=0$. The only positive operator
with zero trace is the zero operator, so it follows that
$E_{yes}({\omega})=0$.

To prove Eq. (7.5), we note that if ${\omega}<1$ then ${\cal H}_{12}$ will have a possibly nonorthogonal basis set
$\{|z_{k}{\rangle}\}$ with the property
\begin{equation}
|{\langle}{\psi}_{1}|z_{k}{\rangle}|={\omega}.
\end{equation}
This condition can always be satisfied for ${\omega}<1$.  For
example, we may take
$|z_{1}{\rangle}={\omega}|{\psi}_{1}{\rangle}+\sqrt{1-{\omega}^{2}}|{\psi}^{\perp}_{1}{\rangle}$
and
$|z_{2}{\rangle}={\omega}|{\psi}_{1}{\rangle}-\sqrt{1-{\omega}^{2}}|{\psi}^{\perp}_{1}{\rangle}$
where ${\langle}{\psi}_{1}|{\psi}^{\perp}_{1}{\rangle}=0$ and
$|{\psi}^{\perp}_{1}{\rangle}$ is the state vector in ${\cal
H}_{12}$ orthogonal to $|{\psi}_{1}{\rangle}$.  Notice that Eq.
(7.10) cannot be satisfied if ${\omega}=1$ since, when this is the
case, the states $|z_{1}{\rangle}$ and $|z_{2}{\rangle}$ will
differ at most by a phase and so they cannot be a basis for ${\cal
H}_{12}$.  In the case of ${\omega}<1$, we may then write
\begin{equation}
|{\psi}_{2}{\rangle}=\sum_{k=1}^{2}c_{k}|z_{k}{\rangle}.
\end{equation}
Suppose now that
$|{\langle}{\psi}_{2}|{\psi}_{1}{\rangle}|{\neq}{\omega}$ and
calculate
\begin{equation}
{\langle}{\psi}_{1}|{\otimes}{\langle}{\psi}_{2}|E_{no}({\omega})|{\psi}_{1}{\rangle}{\otimes}|{\psi}_{2}{\rangle}=\sum_{k,k'=1}^{2}c^{*}_{k'}c_{k}{\langle}{\psi}_{1}|{\otimes}{\langle}z_{k'}|E_{no}({\omega})|{\psi}_{1}{\rangle}{\otimes}|z_{k}{\rangle}.
\end{equation}
\\
As we shall now see, all terms in this sum are vanishing.  To
prove this, we again use the Cauchy-Schwarz inequality
\begin{equation}
|{\langle}{\psi}_{1}|{\otimes}{\langle}z_{k'}|E_{no}({\omega})|{\psi}_{1}{\rangle}{\otimes}|z_{k}{\rangle}|^{2}{\leq}{\langle}{\psi}_{1}|{\otimes}{\langle}z_{k'}|E_{no}({\omega})|{\psi}_{1}{\rangle}{\otimes}|z_{k'}{\rangle}{\langle}{\psi}_{1}|{\otimes}{\langle}z_{k}|E_{no}({\omega})|{\psi}_{1}{\rangle}{\otimes}|z_{k}{\rangle}=0,
\end{equation}
where the equality follows from combining Eq. (7.2) with Eq.
(7.10).  So,
${\langle}{\psi}_{1}|{\otimes}{\langle}{\psi}_{2}|E_{no}({\omega})|{\psi}_{1}{\rangle}{\otimes}|{\psi}_{2}{\rangle}=0$
for all $|{\psi}_{1}{\rangle},|{\psi}_{2}{\rangle}{\in}{\cal H}$.
In the same way as with $E_{yes}({\omega})$, we can use this to
prove that for ${\omega}{\in}[0,1)$, $E_{no}({\omega})=0$.  To do
so, we calculate the trace of $E_{no}({\omega})$ in a product
basis and use Eqs. (7.12) and (7.13) to obtain
$\mathrm{Tr}[E_{no}({\omega})]=0$. This, together with the
positivity of $E_{no}({\omega})$ implies that
$E_{no}({\omega})=0$.

The upshot of this is that, for any given value of the overlap
${\omega}$, it is impossible to unambiguously confirm, always with
non-zero probability, that for two particles prepared in
arbitrary, unknown pure states with overlap ${\omega}$, that
${\omega}$ is indeed the value of their overlap. Furthermore, when
their overlap is not equal to ${\omega}$, it is also impossible to
confirm this, again with non-zero probability for all states,
unless ${\omega}=1$, which corresponds to the case of the two
states being identical. So, the only kind of universal unambiguous
overlap filtering which is possible is universal difference
detection.

\section{Discussion}
\renewcommand{\theequation}{8.\arabic{equation}}
\setcounter{equation}{0} \label{sec:8} In this paper, we have
investigated the possibility of unambiguous state comparison.  As
with many other operations on quantum systems, such as state
discrimination and cloning, the nature of quantum operations and
measurements imposes restrictions upon our ability to carry this
out. We began by examining the possibility of determining when the
states of $N$ particles are all identical.   If we wish to
unambiguously determine when this is the case, and always with
non-zero probability when it is true, then the possible states of
the individual particles must form a linearly independent set.
This result adds to the growing number of probabilistic quantum
operations which are only possible on linearly independent sets.
Among these are unambiguous state discrimination \cite{Chefles1},
probabilistic cloning \cite{DG} and unambiguous discrimination
among unitary operators \cite{retro}.

Although it is impossible to unambiguously confirm identicality if
these single-particle states are linearly dependent, there remains
the possibility of confirming unambiguously that they are not all
identical.  We have referred to this as difference detection and
found that it can be done for all $N$-particle product states with
a single measurement.  This is simply a collective, von Neumann
measurement on the $N$-particle system which distinguishes between
the symmetric subspace and its orthogonal complement, the
asymmetric subspace. We found that it is actually optimal in
situations where all product states of the form
$|{\Psi}{\rangle}=|{\psi}{\rangle}^{{\otimes}N}$ are possible. The
experimental implementation of this measurement is discussed in
\cite{JAC}, where the comparison of photon polarisation states is
considered.  The emphasis is on the realisation of this
measurement using only linear optical elements.  It is shown that
this can be achieved perfectly for a pair of photons.  It is also
shown here that unambiguous difference detection can be
accomplished for more than two photons using linear optical
techniques, although possibly with less than the theoretical
minimum probability of inconclusive results in Eq. (4.24).

We also considered the problem of determining unambiguously when all $N$ particles are in different states.  For this to be possible,
we found that a necessary and sufficient condition is that the actual states of the $N$ particles form a linearly independent set.  We
derived a universal measurement which can be used to confirm that all $N$ particles are in different states, whenever this linear
independence condition is satisfied.  This measurement is a projective measurement which distinguishes between the antisymmetric
subspace an its orthogonal complement, the non-antisymmetric subspace.  We showed that this measurement is optimal when all
$N$-particle product states are possible.

For a particular $N$-particle product state, the conclusive result
probabilities for these universal measurements, given by Eqs.
(6.1) and (6.2), depend on the states of the individual particles.
We considered the possibility that these probabilities can serve
as measures of the distinguishability of these single-particle
states.  We found that they do indeed possess many of the
attributes required of a distinguishability measure.  The
probabilities in Eqs. (6.1) and (6.2) both maximise for orthogonal
states. They also minimise for sets of linearly dependent and
identical states respectively. While identical sets of states are
always considered to have minimum (zero) distinguishability,
linearly dependent sets also have this property in some contexts,
e.g. in unambiguous state discrimination.

One additional requirement of a distinguishability measure is that
it should be non-increasing under a deterministic quantum
operation. We confined our attention to deterministic operations
which transform one set of pure states into another, since all of
our analysis of state comparison is concerned with such states. We
found that the conclusive result probability in Eq. (6.1) is
indeed non-increasing under such transformations.  The situation
is more complicated for the success probability in Eq. (6.2).
While we were able to confirm that it is non-increasing for sets
of two and three states, the question of whether or not this is
true for all $N$ is equivalent to an open conjecture relating to
permanents and Hadamard products of matrices. The investigation
and resolution of this matter is of considerable importance in its
own right.  However, our demonstration that the outcome will have
implications for quantum measurement and information theory gives
a further incentive to pursue it.  As it happens, in the simplest
case of just two states, the non-increasing of these probabilities
is equivalent to the non-decreasing of the overlap of the pair of
states.

We concluded with the exploration of a potential generalisation of
unambiguous state comparison which we termed unambiguous overlap
filtering. When comparing the states of two quantum systems, if
these states are pure, then comparison of these states is
equivalent to determining whether or not their overlap is equal to
1.  This suggests the following generalisation: can we determine,
unambiguously, whether or not the (pure) states of two quantum
systems have overlap equal to ${\omega}$ where ${\omega}$ may take
any fixed value between 0 and 1? What we have found is that it is
impossible to unambiguously confirm that two unknown pure states
have overlap ${\omega}$ for any value of ${\omega}$. This
generalises our earlier result that unambiguous identicality
confirmation for a pair of unknown pure states is impossible
\cite{BCJ} (which is explained in this paper by the fact that
unambiguous identicality confirmation is possible only for
linearly independent sets.) More surprising, however, was our
discovery that unambiguous confirmation that the overlap is not
equal to ${\omega}$ for unknown pure states is possible only if
${\omega}=1$. This, of course, corresponds to difference
detection. So, the only kind of unambiguous overlap filtering for
unknown pure states which is physically possible is unambiguous
difference detection.

\section*{Acknowledgements}

We would like to thank Stephen M. Barnett for interesting
discussions and Ravindra B. Bapat for helpful correspondence. AC
was supported by a UK Engineering and Physical Sciences Research
Council Fellowship in Theoretical Physics and by a University of
Hertfordshire Postdoctoral Research Fellowship. EA was supported
by the EU Marie Curie programme, project number HPMF-CT-2000-00933
and IJ  was supported by the Ministry of Education of the Czech
republic (MSMT 210000018), GACR 202/01/0318.


\begin{thebibliography}{99}
\bibitem{Footnote1}
Throughout this article, when we speak of positive operators and
matrices, we mean operators and matrices that are positive
semidefinite rather than strictly positive definite.
\bibitem{Footnote2}
We take the support of a positive operator to be the subspace
spanned by its eigenstates corresponding to non-zero eigenvalues.
\end{thebibliography}
\end{document}